\documentclass[aps,prb,twocolumn,superscriptaddress,amsmath,amssymb,floatfix]{revtex4-2}

\usepackage{graphicx}
\usepackage{bm}
\usepackage{booktabs}
\usepackage{array}[=2016-10-06]
\usepackage{xcolor}
\usepackage{enumitem}
\usepackage[
    colorlinks=true,
    linkcolor=PastelPurple,
    citecolor=PastelBlue,
    urlcolor=PastelBlue
]{hyperref}

\definecolor{PastelPurple}{HTML}{9D6CA9}
\definecolor{PastelBlue}{HTML}{648EBE}
\usepackage{comment}
\usepackage{tikz}
\usepackage{physics}
\usepackage{orcidlink}
\usepackage{tabularx}

\newcommand{\lamc}{\lambda_{\mathrm c}}
\newcommand{\ER}{E_{\mathrm R}}
\newcommand{\ERz}{E_{\mathrm R,z}}
\newcommand{\Jex}{J_{\mathrm{ex}}}
\newcommand{\dxy}{d_{x^2-y^2}}
\newcommand{\dz}{d_{z^2}}
\newcommand{\Sr}{{}^{87}\mathrm{Sr}}
\newcommand{\rvec}{\bm r}
\newcommand{\kvec}{\bm k}
\newcommand{\Bog}{B_{1g}}
\newcommand{\up}{\uparrow}
\newcommand{\dn}{\downarrow}
\newcommand{\avg}[1]{\overline{#1}}
\newcommand{\Jph}{J_{\mathrm{ph}}}

\usepackage{soul} 

\usepackage[normalem]{ulem} 

\newcommand{\Harvard}{Department of Physics, Harvard University, Cambridge, MA 02138, USA}
\newcommand{\ITAMP}{ITAMP, Center for Astrophysics $\vert$ Harvard \& Smithsonian, Cambridge, MA 02138, USA}
\newcommand{\MIT}{Research Laboratory of Electronics, Massachusetts Institute of Technology, Cambridge, Massachusetts 02139, USA}
\newcommand{\ASTAR}{Institute of Advanced Intelligence and Computing (IAIC), Agency for Science, Technology and Research (A*STAR), Singapore 138632, Republic of Singapore}

\begin{document}

\title{
Quantum simulation of material-specific Hubbard models with alkaline-earth atoms
}

\author{Henning Schl\"omer\orcidlink{0000-0002-3879-9837}}
\thanks{These authors contributed equally.}
\affiliation{\ITAMP}
\affiliation{\Harvard}
\author{Reuben R. W. Wang\orcidlink{0000-0002-1069-9746}}
\thanks{These authors contributed equally.}
\affiliation{\ITAMP}
\affiliation{\Harvard}
\author{Bo Xing~\orcidlink{0000-0002-9651-5505}}
\affiliation{\MIT}
\affiliation{\ASTAR}
\author{Susanne F. Yelin~\orcidlink{0000-0003-1655-9151}}
\affiliation{\Harvard}
\author{H. R. Sadeghpour\orcidlink{0000-0001-5707-8675}}
\affiliation{\ITAMP}

\date{\today}

\begin{abstract}
A long-term goal of quantum simulation is to faithfully reproduce the low-energy Hamiltonians of correlated quantum materials, establishing a direct connection between solid-state experiment, electronic-structure theory, and programmable many-body systems. Here, we introduce an integrated quantum-simulation architecture based on fermionic alkaline-earth atoms that enables programmable control over the key ingredients of multi-orbital material Hamiltonians. Long-lived clock states encode synthetic orbitals, their differential polarizabilities generate orbital-selective tunneling, and contact interactions provide multi-orbital Hubbard and Hund couplings. Spatially structured coherent driving on the ultra-narrow clock transition further realizes symmetry-selective inter-orbital hybridization, while the clock detuning independently controls orbital energy offsets. In this way, the spatial symmetries of solid-state orbitals can be encoded directly into the effective Hamiltonian, even though the atoms themselves remain in conventional $s$-wave Wannier orbitals of the lowest optical-lattice bands. We demonstrate the resulting integrated control using the four-band low-energy Hamiltonian of the bilayer nickelate La$_3$Ni$_2$O$_7$ as a demanding benchmark. Self-consistent calculations show that the microscopic parameters obtained from first-principles modeling can be reproduced within a single experimentally accessible lattice configuration and independently programmed over substantial parameter ranges. Our results establish a general route for translating material-derived multi-orbital Hamiltonians into programmable ultracold-atom quantum simulators.
\end{abstract}

\maketitle

\section{Introduction}

The Fermi-Hubbard model (FHM) is the bedrock model of strongly correlated fermions~\cite{Arovas2022}, capturing essential correlation-driven physics across a broad class of quantum materials~\cite{Scalapino2012}. 
In the past few decades, ultracold fermions in optical lattices have emerged as quantitative simulators for the single-band FHM~\cite{esslinger2010,gross2017,Carroll25_Sci, BohrdtReview}. Combined with quantum gas microscopy~\cite{grossbakr2021}, these platforms have resolved antiferromagnetic correlations~\cite{Greif2013, Hart2015, Boll2016, Mazurenko_AFM}, magnetic polarons~\cite{Koepsell_nature2019, Chiu2019, Hartke2020, Koepsell2021, Ji2021}, as well as stripe~\cite{Bourgund2025}, pairing~\cite{Hirthe2023} and pseudogap physics~\cite{chalopin2024, Kendrick2025}, providing a controlled platform for exploring doped Mott insulators in strongly correlated regimes at progressively lower temperatures~\cite{xu2025}.

\begin{table*}
\caption{Requirements for multi-orbital simulation and their implementation with our alkaline-earth atom architecture.}
\label{tab:toolbox}
\centering

\renewcommand{\arraystretch}{1.3}
\setlength{\tabcolsep}{10pt}
\small

\begin{tabularx}{\textwidth}{
    @{}
    >{\raggedright\arraybackslash}p{0.09\textwidth}
    @{\hspace{10pt}}
    X
    X
    @{}
}
\toprule

&
\textbf{Material requirement}
&
\textbf{Quantum simulator implementation}
\\

\midrule

\vspace{-5pt}
\includegraphics[width=\linewidth]{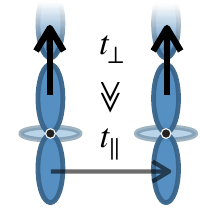}
&
\textit{Hopping hierarchy.}
Tunneling amplitudes are generally orbital-dependent and spatially anisotropic, since each orbital transfers efficiently only along the directions in which its lobes point.
&
\textit{State-dependent optical lattices.}
The differential polarizability of the clock states assigns each synthetic orbital its own tunneling amplitudes. Independent control of the in-plane and vertical confinement realizes anisotropic orbital-dependent hopping hierarchies.
\\

\addlinespace

\vspace{-7pt}
\includegraphics[width=\linewidth]{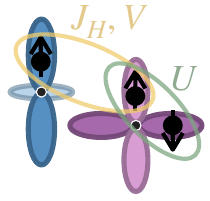}
&
\textit{Interaction structure.}
Interactions in multi-orbital models are characterized by orbital-resolved Hubbard repulsions, direct inter-orbital terms, and intra-atomic Hund exchange.
&
\textit{Contact interactions.}
Interatomic $s$-wave collisions generate orbital-resolved Hubbard interactions, direct inter-orbital terms, and Hund exchange, tunable through the lattice confinement or available Feshbach resonances.
\\

\addlinespace

\vspace{-7pt}
\includegraphics[width=\linewidth]{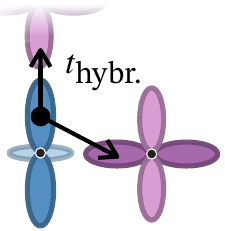}
&
\textit{Hybridization \& crystal field.}
Inter-orbital tunneling is constrained by orbital symmetry and can change sign between crystallographic directions or vanish along symmetry-forbidden bonds. Crystal-field splittings determine the relative orbital energies.
&
\textit{Structured clock coupling.}
A spatially structured field driving the clock transition acts as an optical ``mask'' for coherent orbital coupling. Optical interference patterns generate local amplitudes and phases that imprint the desired inter-orbital tunneling structure. The laser detuning acts as the crystal-field splitting.
\\

\bottomrule
\end{tabularx}

\end{table*}

A practical goal of quantum simulation is to move beyond paradigmatic lattice models and accurately simulate the low-energy Hamiltonians of strongly correlated quantum materials. A key challenge in this direction is that the relevant low-energy degrees of freedom are often distributed across several active orbitals. Iron-based superconductors, for example, are intrinsically multi-orbital, with several Fe $3d$ orbitals contributing near the Fermi level~\cite{raghu2008,fernandes2022}; layered nickelates host multiple active Ni $3d$ orbitals that are central to their low-energy electronic structure~\cite{puphal2026}; and in Hund metals such as the ruthenates, multiple orbitals with orbital-dependent hopping, hybridization, and Hund exchange govern their correlated electronic structure~\cite{georges2013}. 

\begin{figure}[th!]
\includegraphics[width=\columnwidth]{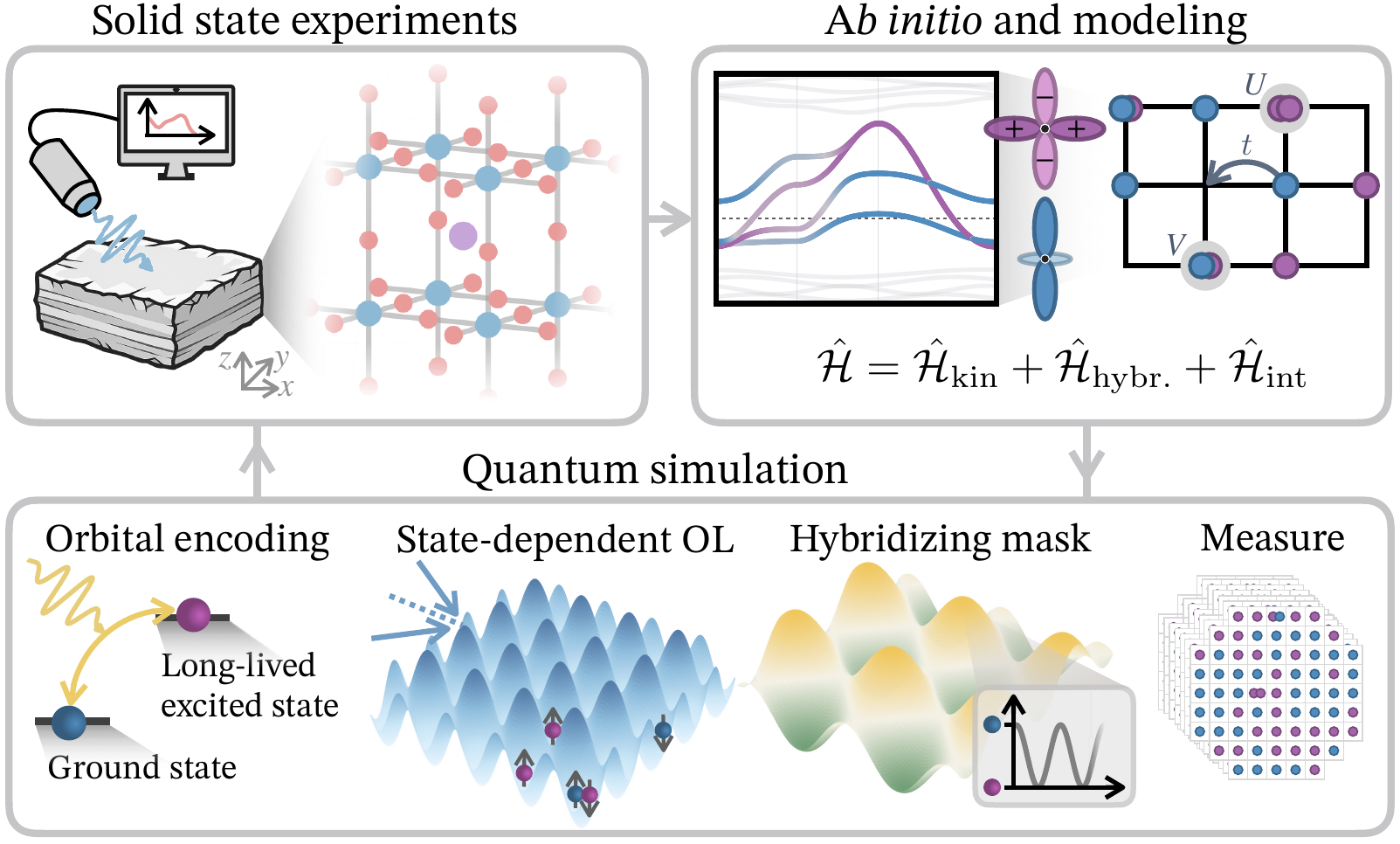}
\caption{\textbf{Toolbox for material-realistic quantum simulation.} Lower panel: a quantum simulator that combines orbital encoding into long-lived clocks states of Sr,
orbital-dependent optical potentials, spatially structured coherent coupling for symmetry-selective hybridization, and snapshot measurements. These ingredients enable the implementation and controlled exploration of multi-orbital Hamiltonians derived from electronic-structure calculations. Such a platform closes a loop between solid-state experiments (upper left panel), first-principles modeling (upper right panel), and quantum simulation, in which experimentally constrained microscopic models are solved on a programmable quantum simulator and the resulting insights feed back into the understanding and design of correlated materials.}
\label{fig:loop}
\end{figure}

Implementation of such material-realistic models would place quantum simulators in direct contact with experiment and first-principles electronic-structure theory, closing the loop sketched in Fig.~\ref{fig:loop}. Materials measurements constrain microscopic structure, electronic-structure calculations~\cite{DFT} reduce this information to an effective low-energy Hamiltonian, and a quantum simulator solves the resulting model in regimes inaccessible to controlled classical methods. 

Reaching this level of materials realism, however, imposes substantial demands on the simulator. Solid state orbitals are distinguished not only by their energies and interactions, but also by their spatial symmetry. Their directional lobes and parity determine which hopping and hybridization processes are allowed, their relative magnitudes, and their signs. These structures are not inherently native to ultracold atoms. When loaded into the lowest band of a conventional optical lattice, ultracold atoms share $s$-wave Wannier envelopes; {\it the orbital structure of the material Hamiltonian must therefore be 
engineered into the quantum simulator.} More specifically, a programmable simulator of material-realistic multi-orbital models must provide control over recurring ingredients of these systems, including orbital-dependent hopping, intra- and inter-orbital interactions, and inter-orbital hybridization, as listed in Tab.~\ref{tab:toolbox}. 

Recent proposals and experiments have begun to move beyond the single-band Hubbard paradigm. One route encodes orbital degrees of freedom in the geometry of the optical potential; for example, Lieb lattices have enabled multi-orbital structures~\cite{LebratLiebLattice} and provide a route toward Emery-type charge-transfer models of the cuprates~\cite{Homeier2026,Lange2026}. Alternatively, internal atomic states can encode distinct orbitals. In alkaline-earth atoms (AEA), the long-lived ground and excited clock states provide two such states~\cite{gorshkov2010}, motivating orbital-dependent Hubbard models~\cite{Sotnikov2020,Bilokon2023} and enabling SU($N$) magnetism~\cite{Ozawa2018,Taie2022,Li2026_preprint}. Neither potential shaping nor internal-state encoding alone, however, provides independent control over the full hopping, interaction, and hybridization structure.

We propose a quantum simulator based on fermionic alkaline-earth atoms that combines these material-specific requirements within a single microscopic architecture, summarized in Fig.~\ref{fig:loop}. Two long-lived clock states encode synthetic orbitals, while orbital-selective optical potentials control their hopping and spatial anisotropy, and contact interactions generate the required intra-orbital, inter-orbital, and exchange couplings. A spatially structured field resonant with the ultranarrow clock transition acts as a coherent-coupling mask, with its amplitude and phase imprinting the desired magnitude and sign structure of the inter-orbital hybridization. The clock detuning controls the crystal-field splitting. In this way, the symmetry of the material orbitals is encoded in matrix elements between locally $s$-like Wannier states. Table~\ref{tab:toolbox} summarizes the correspondence between the recurring requirements of multi-orbital material Hamiltonians and the controls available in the proposed AEA architecture. 

\begin{figure*}[ht]
\includegraphics[width=\textwidth]{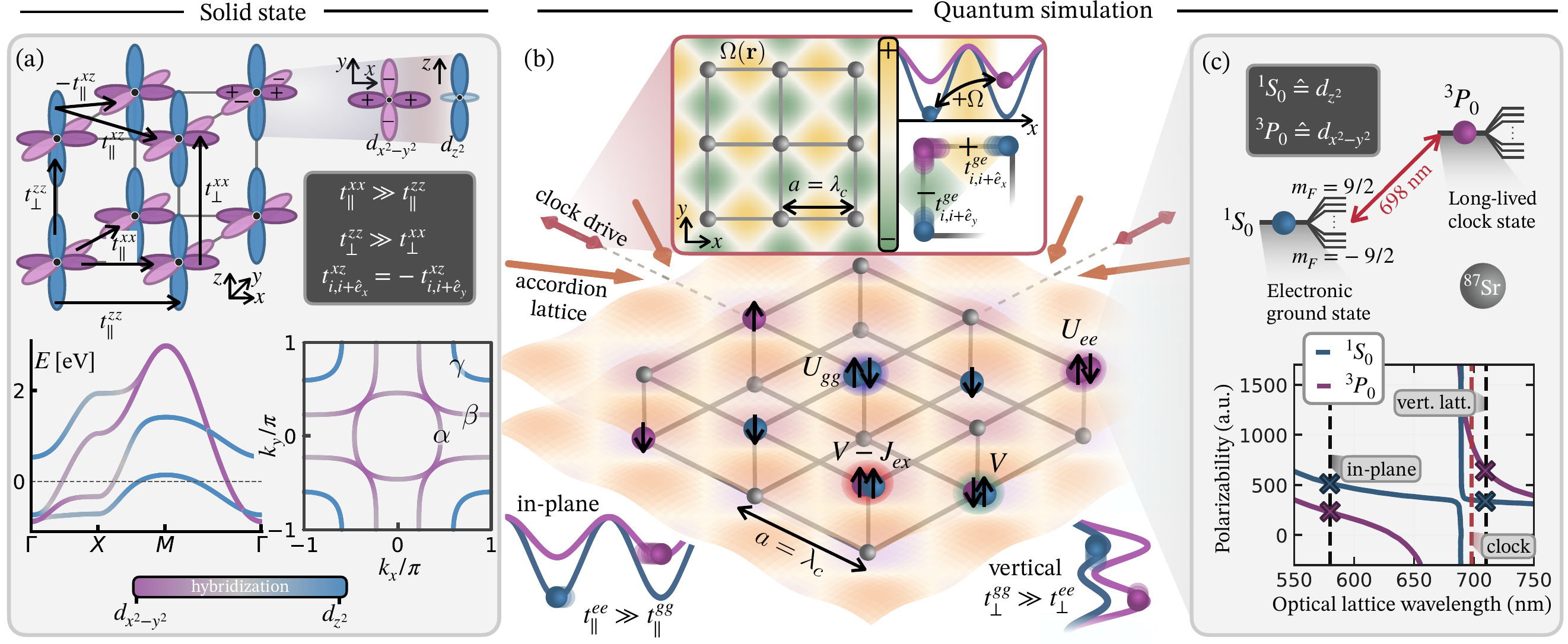}
\caption{\textbf{Recipe for multi-orbital quantum simulation of the bilayer nickelate.} (a) La$_3$Ni$_2$O$_7$ under pressure realizes a tetragonal bilayer of Ni atoms, each contributing two $e_g$ orbitals, $\dxy$ and $\dz$. The orbital symmetry dictates the hopping hierarchy ($t^{xx}_\parallel \gg t^{zz}_\parallel$ in plane, $t^{zz}_\perp \gg t^{xx}_\perp$ across the bilayer) and a $d$-wave inter-orbital hybridization $t^{xz}_{i,i+\hat{e}_x} = -t^{xz}_{i,i+\hat{e}_y}$, producing the (non-interacting) band structure and $\alpha$, $\beta$, $\gamma$ Fermi pockets shown. (b) $\Sr$ in an accordion lattice of spacing $a=\lamc$ (the clock transition wavelength) forms the in-plane square lattice; a vertical state-dependent superlattice forms the bilayer. On-site interactions $U_{gg}, U_{ee}$, direct $V$, and Hund exchange $\Jex$ arise from collisions. An optical lattice at the clock transition imprints the $d$-wave Rabi mask $\Omega(\rvec)$ (top inset), which vanishes on every site and alternates sign between $x$ and $y$ bonds, realizing the target hybridization. (c) Level scheme with the orbital mapping $e={}^3P_0 \leftrightarrow \dxy$, $g={}^1S_0 \leftrightarrow \dz$, together with the polarizabilities of $g$ and $e$. The electronic spin degree of freedom is encoded into the atomic hyperfine sublevels $m_F$.}
\label{fig:overview}
\end{figure*}

We demonstrate this integrated control using the bilayer nickelate La$_3$Ni$_2$O$_7$ as a demanding concrete target in which all three classes of requirements in Table~\ref{tab:toolbox} define its low-energy physics. Following the discovery of superconductivity at $T_c \approx 80$~K under pressure~\cite{sun2023}, La$_3$Ni$_2$O$_7$ has drawn considerable interest as the only other family of correlated materials, after cuprates, to superconduct above the boiling point of liquid nitrogen. This makes La$_3$Ni$_2$O$_7$ an especially compelling test case for quantum simulation, where correlations and multiple active orbitals must be treated on equal footing to address the underlying many-body problem.

More broadly, while La$_3$Ni$_2$O$_7$ provides the central example considered in this work, the proposed architecture is not specific to the nickelate Hamiltonian. It should instead be viewed as a general blueprint for the quantum simulation of hybridized multi-orbital Hubbard models, providing internal-state encoding of orbitals, orbital-selective trapping, spatially structured coherent coupling, and tunable orbital energies.

The remainder of the paper is organized as follows. Sec.~\ref{sec:model} introduces the target nickelate model and collects the single-particle and interaction parameters that an accurate quantum simulator must reproduce. Sec.~\ref{sec:summary} summarizes the recipe of our toolbox, presenting the optimized Hamiltonian parameters of the quantum simulator. Sec.~\ref{sec:implementation} develops the technical details of the alkaline-earth system. Finally, in Sec.~\ref{sec:conclusions}, we conclude.

\section{Bilayer nickelate model}
\label{sec:model}

We start by introducing the interacting model of bilayer nickelates under pressure, before using our proposed toolbox to build a concrete implementation protocol in the proceeding sections.

Early \textit{ab initio} calculations of La$_3$Ni$_2$O$_7$ have isolated two orbitals per Ni atom as the active degrees of freedom at low energies, allowing a four-band per bilayer cell model to accurately capture the low-energy physics~\cite{luo2023}. While calculations in certain limits such as weak coupling are possible, the enlarged local Hilbert space places the correlated regime beyond controlled classical methods, and predictions for the pairing symmetry remain unsettled~\cite{yang2023,liu2023,lu2024,sakakibara2024}. This is precisely where quantum simulators can provide genuinely new insights, provided they accurately reproduce the orbital and interaction structure of the material model. Simplified single-band models of bilayer nickelates have been addressed for ultracold atom~\cite{Schloemer_proposal} and trapped ion~\cite{Granet2026} platforms, but the simulation of more material-realistic models remains a significant open challenge.

\subsection{Single-particle Hamiltonian}
\label{subsec:target_single_particle}
Under pressure, bilayer nickelates form a tetragonal bilayer of NiO$_2$ planes~\cite{sun2023}, with two orbitals per Ni atom and two Ni atoms per bilayer cell, see Fig.~\ref{fig:overview}~(a). The two active orbital degrees of freedom are partially filled $\dxy$ and $\dz$ orbitals. In the basis $\{x \equiv \dxy,\ z \equiv \dz\}$, and with $\ell=1,2$ labeling the two layers, the (non-interacting) tight-binding Hamiltonian reads~\cite{luo2023}
\begin{align}
\hat{\mathcal{H}}_0 ={}& -\sum_{\langle ij\rangle\ell\gamma\sigma} t^{\gamma\gamma}_{\parallel}\,
\hat{c}^\dagger_{i\ell\gamma\sigma} \hat{c}_{j\ell\gamma\sigma} - \sum_{i\gamma \sigma} t_\perp^{\gamma \gamma} \hat{c}^\dagger_{i1\gamma \sigma}\hat{c}_{i2 \gamma \sigma} + \mathrm{h.c.}   \nonumber\\
&- \sum_{\langle ij\rangle\ell\sigma} t^{xz}_{ij}\big(
\hat{c}^\dagger_{i\ell x\sigma} \hat{c}_{j\ell z\sigma} + \mathrm{h.c.}\big) \label{eq:nickelate} \\
&+ \frac{\Delta}{2}\sum_{i\ell\sigma}\big(\hat{n}_{i\ell x\sigma}-\hat{n}_{i\ell z\sigma}\big), \nonumber
\end{align}
where $\gamma=x,z$ labels orbitals and $\sigma=\uparrow,\downarrow$ denotes spin. 

The first line describes tunneling within and between NiO$_2$ planes: $t^{\gamma\gamma}_\parallel$ is the intra-orbital nearest-neighbor hopping of orbital $\gamma$ between in-plane sites $i$ and $j$ of the same layer $\ell$, while $t_\perp^{\gamma\gamma}$ tunnel couples the two layers on the same in-plane site $i$. The second line is the hybridization term: $t^{xz}_{ij}$ transfers an atom between the two orbitals on neighboring in-plane sites and is the only term that breaks the orbital number conservation. There is no vertical inter-orbital nearest-neighbor term, since $\dxy$ and $\dz$ transform differently under the point-group operations that leave the vertical bond invariant, so their hybridization along $z$ vanishes by symmetry. Likewise, on-site hybridization is forbidden because the same rotation distinguishes the two orbitals, so any on-site potential respecting the tetragonal symmetry can only split them, but not mix them. This splitting from the crystal field $\Delta$ is the last term, which acts as a static orbital field and fixes the relative filling of the two orbitals. Note that we have omitted longer-range hopping, such as in-plane and out-of-plane diagonal couplings, as they are negligible compared to the dominant hopping scales~\cite{luo2023}.

Three features encode the orbital symmetry, illustrated pictorially in Fig.~\ref{fig:overview}~(a). First, the in-plane tunneling is strongly orbital selective: the planar $\dxy$ orbital disperses much more than $\dz$, so $t^{xx}_\parallel \gg t^{zz}_\parallel$. Second, the hierarchy inverts across the bilayer: the axially oriented $\dz$ orbital dominates the interlayer coupling, so $t^{zz}_\perp \gg t^{xx}_\perp$. Third, the inter-orbital hybridization inherits the form factor of the $\dxy$ orbitals. It has equal magnitude but opposite sign on $x$ and $y$ bonds,
\begin{equation}
 t^{xz}_{i,i+\hat e_x}=+t^{xz}_{\parallel},
 \qquad
 t^{xz}_{i,i+\hat e_y}=-t^{xz}_{\parallel},
 \label{eq:d_wave_hopping}
\end{equation}
corresponding to the $\Bog$ symmetry $(\cos k_x - \cos k_y)$. 

Hopping values as obtained from \textit{ab initio} calculations~\cite{luo2023} are summarized in Table~\ref{tab:hoppings}. Diagonalizing Eq.~\eqref{eq:nickelate} gives two electron pockets $\alpha,\beta$ and a hole pocket $\gamma$, illustrated in Fig.~\ref{fig:overview}~(a). The $\gamma$ pocket is almost entirely composed of $\dz$ orbitals, and its size is controlled by the $\dz$ interlayer bonding-antibonding splitting, which in turn is tied to the crystal field $\Delta$. The $\alpha$ and $\beta$ sheets, on the other hand, hybridize between both orbitals, with maximal mixing at the zone boundaries. While different DFT calculations can yield slightly different quantitative values, the qualitative hierarchy discussed above remains the same. 
We will show that appropriate engineering of an AEA simulator can match the non-interacting band structure tabulated in Table~\ref{tab:hoppings} with high precision.   

\begin{table}[t]
\caption{Single-particle parameters of the bilayer nickelate model~\cite{luo2023}, given in eV and normalized to the in-plane hopping $t^{xx}_\parallel$.}
\label{tab:hoppings}
\begin{ruledtabular}
\begin{tabular}{lcccccc}
 & $t^{xx}_\parallel$ & $t^{zz}_\parallel$ & $t^{xz}_\parallel$
 & $t^{zz}_\perp$ & $t^{xx}_\perp$ & $\Delta$ \\
\colrule
eV & $0.483$ & $0.110$ & $0.239$ & $0.635$ & $0.005$ & $0.367$ \\
$/t^{xx}_\parallel$ & $1$ & $0.228$ & $0.495$ & $1.315$ & $0.010$ & $0.760$ \\
\end{tabular}
\end{ruledtabular}
\end{table}

\subsection{Interaction structure}
\label{subsec:target_interactions}

Electronic interactions on top of the single-particle Hamiltonian are usually modeled through on-site Hubbard-Kanamori terms~\cite{luo2023,liu2023},
\begin{equation}
\begin{aligned}
\hat{\mathcal{H}}_{\rm int} ={}& \sum_{i\ell\gamma} U_{\gamma}\, \hat{n}_{i\ell\gamma\up} \hat{n}_{i\ell\gamma\dn} + V \sum_{i\ell\sigma\sigma'} \hat{n}_{i\ell x\sigma} \hat{n}_{i\ell z\sigma'} \\
&+ \Jex\sum_{i\ell\sigma\sigma'} \hat{c}^\dagger_{i\ell x\sigma}\hat{c}^\dagger_{i\ell z\sigma'} \hat{c}_{i\ell x\sigma'}\hat{c}_{i\ell z\sigma} \\
&+ \Jph\sum_{i\ell}\big( \hat{c}^\dagger_{i\ell x\up}\hat{c}^\dagger_{i\ell x\dn} \hat{c}_{i\ell z\dn}\hat{c}_{i\ell z\up} + \mathrm{h.c.}\big),
\label{eq:Hint}
\end{aligned}
\end{equation}
with intra-orbital (inter-orbital) direct $U_\gamma$ ($V$)  repulsion, inter-orbital exchange $\Jex$, and pair hopping $\Jph$. If the two orbitals were degenerate and the interaction were invariant under rotations in orbital space, the four amplitudes would not be independent but obey the Kanamori relations
\begin{equation}
\begin{aligned}
    U_x &= U_z \equiv U, \\ 
    V &= U - 2J_H, \\ 
    \Jex &= \Jph = J_H ,
    \label{eq:kanamori}
\end{aligned}
\end{equation}
leaving a single interaction scale $U$ and the Hund coupling $J_H$ to fix the other parameters. This is the form adopted in model calculations for La$_3$Ni$_2$O$_7$, and we use it as the reference point below.

The exchange term becomes more transparent in terms of the orbital-resolved spin operators $\hat{\mathbf{S}}_{i\ell\gamma}$. Reordering the fermion operators and using $\sum_{\sigma\sigma'}\hat{c}^\dagger_{x\sigma}\hat{c}_{x\sigma'} \hat{c}^\dagger_{z\sigma'}\hat{c}_{z\sigma} = 2\,\hat{\mathbf{S}}_{x}\cdot\hat{\mathbf{S}}_{z} + \tfrac{1}{2}\hat n_x \hat n_z$, Eq.~\eqref{eq:Hint} is equivalent to
\begin{align}
    \hat{\mathcal H}_{\rm int} ={}& \sum_{i\ell\gamma} U_{\gamma}\,\hat n_{i\ell\gamma\up}\hat n_{i\ell\gamma\dn} +\Big(V-\frac{\Jex}{2}\Big)\sum_{i\ell}\hat n_{i\ell x}\hat n_{i\ell z} \nonumber\\
    &-2\Jex\sum_{i\ell}\hat{\mathbf{S}}_{i\ell x}\cdot\hat{\mathbf{S}}_{i\ell z} \label{eq:spin_form} \\ &+ \Jph\sum_{i\ell}\big( \hat{c}^\dagger_{i\ell x\up}\hat{c}^\dagger_{i\ell x\dn} \hat{c}_{i\ell z\dn}\hat{c}_{i\ell z\up} + \mathrm{h.c.}\big), \nonumber
\end{align}
showing that $\Jex > 0$ corresponds to a ferromagnetic Hund coupling.

Constrained random phase approximation (cRPA) downfoldings onto the four-orbital subspace estimate the following ranges for the interaction parameters~\cite{christiansson2023},
\begin{equation}
\frac{U}{t^{xx}_\parallel}\approx 6\text{-}8, \qquad \frac{V}{t^{xx}_\parallel}\approx 4\text{-}5, \qquad \frac{J_H}{t^{xx}_\parallel}\approx 1.2\text{-}1.3.
\label{eq:target_interactions}
\end{equation}
We note that the Kanamori relations of Eq.~\eqref{eq:kanamori} are not exactly satisfied, but they hold approximately~\footnote{The two intra-orbital repulsions are nearly identical, $U_x/U_z \approx 1.06$, and $U - 2J_H$ agrees with $V$ to within about $10\%$}.
The essential feature is the hierarchy $U > V > J_H > 0$ with $U$ comparable to the full four-band bandwidth, placing the material in the strongly correlated regime. In our proposal, we show that AEA, specifically $\Sr$, can indeed realize this interaction structure with the correct ordering of interaction strengths; only the pair-hopping term is not generated by contact scattering as discussed in Sec.~\ref{sec:int_AEA}.

\subsection{Open questions and simulation targets}
\label{subsec:open_questions}

The correlated regime of the bilayer nickelate remains theoretically unsettled. The interaction scale is comparable to the full bandwidth, placing the experimentally relevant system between limits in which controlled weak- and strong-coupling methods are available. Although superconducting tendencies are found in both regimes, the resulting low-energy descriptions differ in important respects. This leaves several concrete questions that a quantum simulator of the microscopic Hamiltonian could address: (i) what is the dominant pairing channel, given that different theoretical approaches predict competing superconducting states and pairing symmetries; (ii) what role does the $\gamma$ pocket play, whose predominantly $d_{z^2}$ character and sensitivity to the crystal-field splitting make it a natural tuning knob, while existing theories disagree on whether it is essential for superconductivity or largely incidental~\cite{Qiu2026}; (iii) how do magnetic and pairing correlations evolve with interaction strength across the intermediate-coupling regime; and (iv) how important is the Hund sector, whose sizable inter-orbital exchange provides an additional energy scale absent from single-band descriptions?

Addressing these questions from a quantum simulation point of view requires reproducing the orbital-dependent hopping hierarchy and symmetry-selective hybridization of the material while providing independent control over the correlation strength, inter-orbital exchange, and crystal field. 

\section{Integrated simulator architecture}
\label{sec:summary}

Before developing the individual ingredients of the implementation in detail in Sec.~\ref{sec:implementation}, we summarize the central result of our construction. We find that the material-realistic four-band Hamiltonian of the bilayer nickelate can be implemented within a comparatively simple experimental architecture based on two spatially locked optical patterns in the plane. The first is a state-dependent optical lattice that traps the atoms and controls orbital-dependent tunneling, see Fig.~\ref{fig:overview}~(b). The second is a standing-wave field resonant with the clock transition, which, rather than providing the primary confinement, acts as a coherent coupling between the two synthetic orbitals encoded in the long-lived clock states (which we refer to as $g$ and $e$) of AEA and hybridizes them. 

Crucially, this coherent coupling is spatially structured, illustrated in the inset of Fig.~\ref{fig:overview}~(b). After projection onto the lattice, the clock field generates local inter-orbital matrix elements $\Omega(\mathbf r)$ between the lowest-band Wannier states $w_g$ and $w_e$ of the two clock-states,
\begin{equation}
t_{ij}^{eg}
\propto
\int d^3r\,
w_e^*(\mathbf r-\mathbf R_i)\,
\Omega(\mathbf r)\,
w_g(\mathbf r-\mathbf R_j),
\label{eq:rabi_projection_summary}
\end{equation}
where $\mathbf R_i$ and $\mathbf R_j$ denote lattice-site positions. The spatial amplitude and phase of the optical field are therefore directly transferred to the magnitude and sign of the inter-orbital tunneling. Choosing the field such that the lattice sites coincide with its nodes suppresses on-site orbital conversion, while opposite signs on neighboring $x$ and $y$ bonds generate the $B_{1g}$, or ``$d$-wave,'' hybridization required by Eq.~\eqref{eq:d_wave_hopping}. We refer to this spatially structured clock coupling as a \emph{Rabi mask}.

\begin{figure}[t]
\centering
\includegraphics[width=0.96\columnwidth]{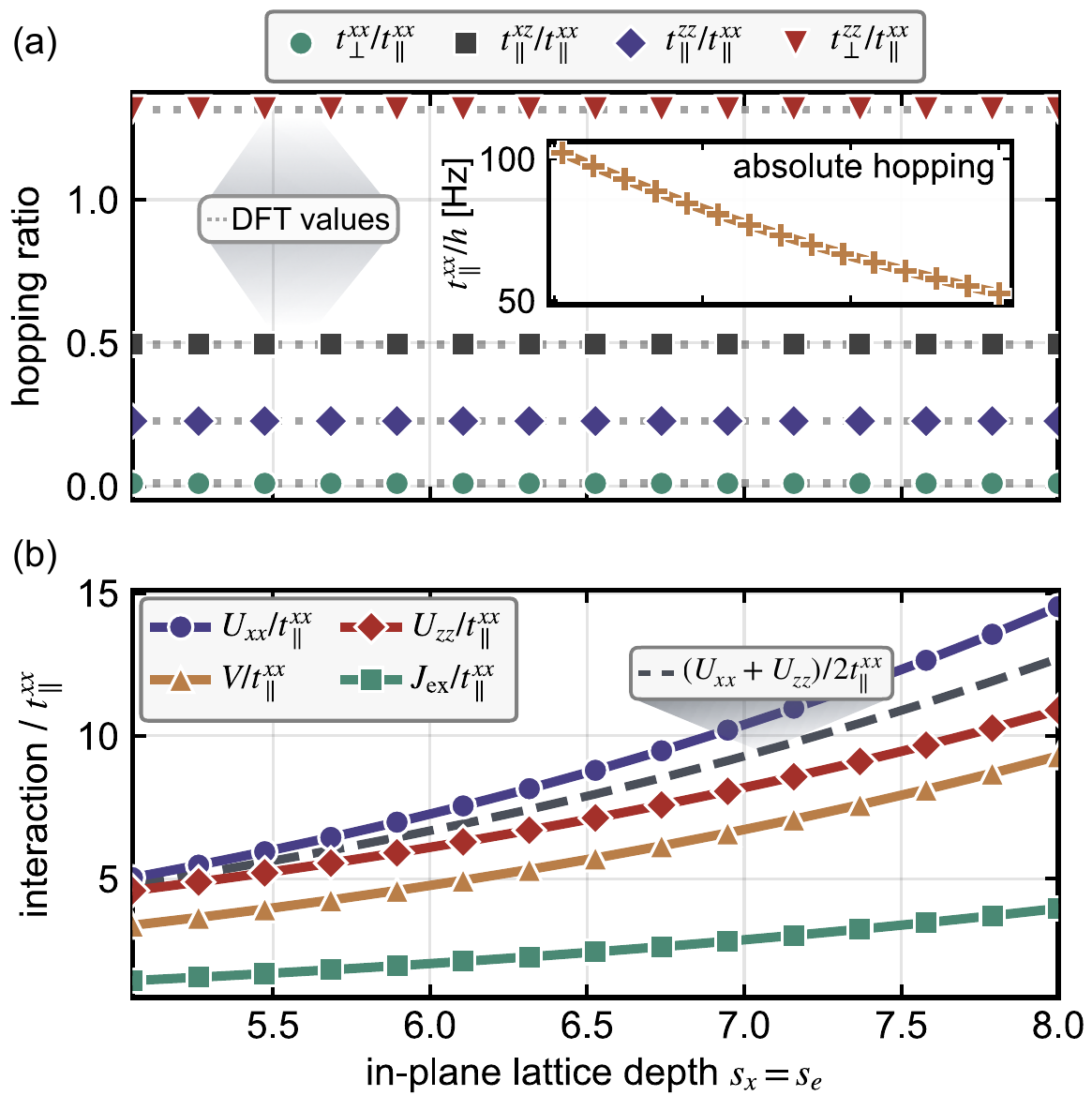}
\caption{\textbf{Accessible parameter regimes of the bilayer-nickelate simulator.}
(a) Optimized hopping amplitudes, normalized to the dominant in-plane hopping $t_{\parallel}^{xx}$, as a function of the in-plane lattice depth (in units of the recoil energy of the clock wavelength). At each depth, the trapping wavelength, clock-drive strength, and vertical superlattice depths are reoptimized self-consistently such that the target hopping structure (Tab.~\ref{tab:hoppings}) is maintained. Gray dotted lines indicate the target values from density functional theory; the inset shows the corresponding absolute hopping scale $t_{\parallel}^{xx}/h$.
(b) On-site interaction parameters obtained from the converged three-dimensional Wannier orbitals. Varying the lattice depth changes the interaction strengths relative to $t_{\parallel}^{xx}$ while preserving the hierarchy between intra-orbital interactions, inter-orbital repulsion, and ferromagnetic Hund exchange. The simulator can therefore access different correlation regimes while keeping the underlying single-particle Hamiltonian essentially fixed.}
\label{fig:optimization_summary}
\end{figure}

While the atoms themselves remain in conventional $s$-like Wannier orbitals of the lowest optical-lattice band, the spatial symmetry of the material orbitals is imprinted onto the effective Hamiltonian through the spatially structured clock field
and the state-dependent trapping potentials. Furthermore, the detuning of the Rabi mask controls the relative energy of the two orbitals, providing direct control over the crystal-field splitting of the multi-orbital Hubbard model. A vertical state-dependent superlattice completes the construction by forming isolated bilayers and generating the inverted orbital hierarchy required for interlayer tunneling.

These controls are not fully independent at the microscopic level. In particular, the clock field that generates the coherent hybridization also induces an off-resonant Stark potential, thereby modifying the Wannier orbitals and the ordinary tunneling amplitudes. We therefore determine the experimental parameters self-consistently. For a chosen in-plane lattice depth, we optimize the trapping wavelength $\lambda_L$ and the clock Rabi frequency such that the in-plane hopping ratio and inter-orbital hybridization simultaneously reproduce their material target values. The vertical superlattice depths are then adjusted to match the two interlayer hopping ratios. At each step, the Wannier functions are recalculated in the full optical potential, ensuring that all hopping, hybridization, and interaction matrix elements are obtained from a single common microscopic lattice configuration. Details of this construction are developed in Sec.~\ref{sec:implementation}.

The resulting optimization is summarized in Fig.~\ref{fig:optimization_summary}. Figure~\ref{fig:optimization_summary}~(a) shows that the target hopping structure from Tab.~\ref{tab:hoppings}, including the clock-induced hybridization, can be maintained over a continuous range of in-plane lattice depths. By readjusting the experimental parameters, the lattice depth can therefore be varied while the single-particle Hamiltonian remains essentially unchanged, allowing the localization of the Wannier orbitals to be tuned independently. As a consequence, the interaction strengths can be varied, as shown in Fig.~\ref{fig:optimization_summary}~(b), while preserving the desired hierarchy between intra-orbital repulsion, inter-orbital repulsion, and ferromagnetic Hund exchange.

These capabilities directly address the open questions identified in Sec.~\ref{subsec:open_questions}. The detuning of the clock drive provides an independent control of the crystal-field splitting and thereby of the occupation of the $d_{z^2}$ orbital and the associated $\gamma$ pocket, offering a direct way to test its role in the correlated state. At the same time, the interaction strength can be varied over a broad range while preserving the underlying fermiology, allowing the evolution of magnetic and pairing correlations across different coupling regimes to be studied without simultaneously changing the band structure. Finally, the nuclear-spin encoding provides an additional handle on the Hund sector, which we develop in Sec.~\ref{subsec:spin_encoding}. The simulator can therefore selectively vary several of the microscopic ingredients whose respective roles in the superconducting and magnetic properties of the bilayer nickelate remain theoretically unresolved.

\section{Implementation with alkaline-earth atoms}
\label{sec:implementation}
We now develop in detail how each term in the bilayer nickelate Hamiltonian can be realized on an ultracold AEA quantum simulator. Since Eq.~\eqref{eq:nickelate} contains two orbitals on each physical lattice site, the first step is to encode this orbital degree of freedom in two long-lived internal atomic states. AEA provides a natural platform for this purpose. Their closed-shell electronic structure gives rise to metastable excited states with exceptionally narrow optical transitions and correspondingly long lifetimes.

We adopt fermionic $\Sr$ as our choice of AEA, as it has favorable contact interaction properties that we will use later. In $\Sr$, we identify the electronic ground state $g={}^1S_0$ and the excited clock state $e={}^3P_0$ with the two synthetic orbitals. The $e={}^3P_0$ state is extremely long-lived, with a lifetime $\tau\gtrsim120~\mathrm{s}$ \cite{Kim25_PRL}, making it well suited to encode an orbital degree of freedom on the timescales relevant for many-body dynamics. The nominally forbidden $g\leftrightarrow e$ transition becomes weakly allowed in fermionic $\Sr$ through hyperfine mixing \cite{Boyd07_PRA}, which can be exploited to coherently couple the two states using light near the clock wavelength $\lambda_c\simeq698.45~\mathrm{nm}$. The remaining spin degree of freedom can then be encoded independently into two nuclear-spin states.

Stated explicitly, we propose the simulator-to-material mapping
\begin{equation} \label{eq:swapped_mapping}
\begin{aligned}
     e &= {}^{3}P_0 \quad\leftrightarrow\quad x=\dxy, \\
     g &= {}^{1}S_0 \quad\leftrightarrow\quad z=\dz,
\end{aligned}
\end{equation} 
as summarized in Fig.~\ref{fig:overview}~(c). Note that the opposite convention, $g\leftrightarrow x$ and $e\leftrightarrow z$, is also feasible at the single-particle level but is more demanding in optical depth and produces a much larger asymmetry between the two intra-orbital repulsions $U_x$ and $U_z$. Such details are developed in Sec.~\ref{subsec:self_consistent} and Appendix~\ref{app:alternative_mapping}.

\subsection{Orbital-selective tunneling}
\label{subsec:hopping}

We first show how the kinetic Hamiltonian terms,
\begin{equation}
\begin{aligned}
\hat{\mathcal{H}}_{\text{kin}} ={} -\sum_{\langle ij\rangle\ell\gamma\sigma} &t^{\gamma\gamma}_{\parallel}\,
\big( \hat{c}^\dagger_{i\ell\gamma\sigma} \hat{c}_{j\ell\gamma\sigma} + \text{h.c.} \big) \\ &- \sum_{i\gamma \sigma} t_\perp^{\gamma \gamma} \big( \hat{c}^\dagger_{i1\gamma \sigma}\hat{c}_{i2 \gamma \sigma} + \text{h.c.} \big),
\label{eq:Hkin} 
\end{aligned}
\end{equation}
are implemented with the hopping hierarchies $t^{xx}_\parallel \gg t^{zz}_\parallel$ and $t^{zz}_\perp \gg t^{xx}_\perp$.

For far-detuned light of intensity $I$, the scalar ac Stark shift $s_\gamma$ experienced by an atom in electronic state $\gamma\in\{g,e\}$ is proportional to its dynamic polarizability $\alpha_\gamma(\lambda)$. Consequently, a single optical lattice of wavelength $\lambda_L$ generally produces different lattice depths for the two clock states, with
\begin{equation}
    \frac{s_e}{s_g}
    =
    \frac{\alpha_e(\lambda_L)}{\alpha_g(\lambda_L)}.
    \label{eq:depth_ratio}
\end{equation}
Away from magic wavelengths, where the two polarizabilities coincide, the states experience different confinement strengths and acquire different tunneling amplitudes. This provides a direct route to orbital-dependent hopping when $g$ and $e$ are used to encode the two synthetic orbitals.

State-dependent lattices of this type have been proposed extensively as a means of realizing Hubbard models with orbital-dependent kinetic energies~\cite{gorshkov2010, Bilokon2023, Sotnikov2020}. 
Here, we use the same mechanism in a targeted way: by explicitly optimizing the lattice wavelength, and thus the state-dependent polarizability ratio, we quantitatively match the orbital-dependent hopping ratios obtained from microscopic electronic-structure calculations for the bilayer nickelate.

\begin{figure}[t]
\centering
\includegraphics[width=0.77\columnwidth]{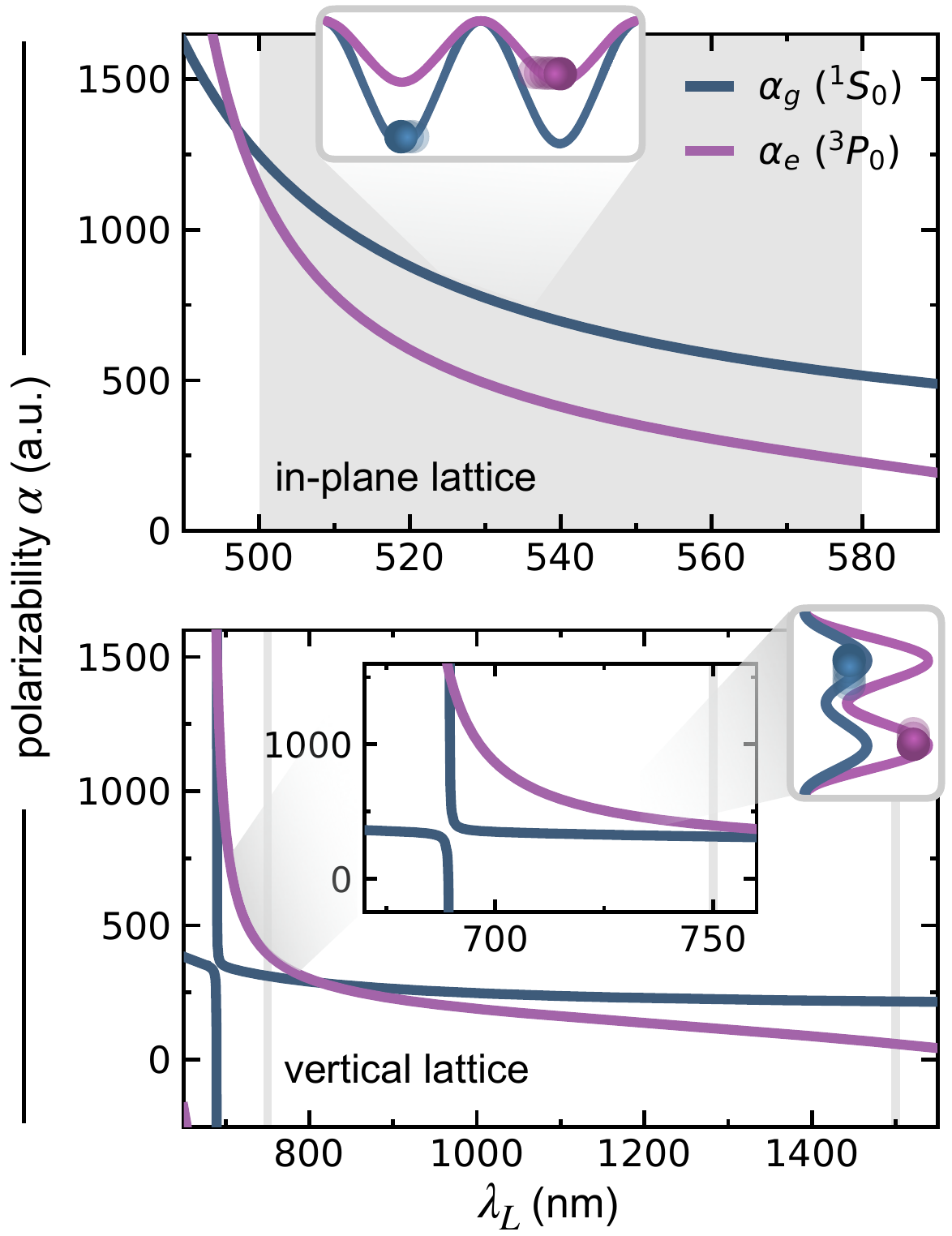}
\caption{\textbf{State-dependent polarizability}. The upper panel shows the in-plane wavelength range in which the $e$ state is trapped more weakly than the $g$ state. The lower panel shows the vertical-lattice wavelengths used in the bilayer construction. The gray bands denote the optimization windows in our simulations.}
\label{fig:polarizability}
\end{figure}

The wavelength-dependent polarizabilities of the two relevant clock states in $\Sr$ are shown in Fig.~\ref{fig:polarizability}, using data from the ATOM database~\cite{safranova2026}. We first consider the interval between $500$ and $580$~nm, highlighted in Fig.~\ref{fig:polarizability} (upper panel). Throughout this range, $\alpha_e/\alpha_g<1$, so the $e$ state experiences a shallower lattice than the $g$ state and consequently acquires a larger tunneling amplitude. With the orbital assignment of Eq.~\eqref{eq:swapped_mapping}, choosing the in-plane lattice wavelength $\lambda_L$ within this window therefore realizes the required hierarchy $t^{xx}_{\parallel}=t^{ee}_{\parallel}>t^{gg}_{\parallel}=t^{zz}_{\parallel}$.

Moreover, because the polarizability ratio varies continuously with wavelength, $\lambda_L$ provides a direct tuning knob for the ratio $t^{ee}_{\parallel}/t^{gg}_{\parallel}$. As discussed in more detail in Sec.~\ref{subsec:self_consistent}, we therefore treat $\lambda_L$ as a free optimization parameter at a fixed in-plane lattice depth to match the target hopping ratio.

The vertical direction requires the opposite hierarchy: the $z$ orbital should tunnel strongly between the two layers of a bilayer, while the corresponding $x$-orbital tunneling should remain small. To realize an isolated bilayer, we form a state-dependent superlattice from short- and long-wavelength components $\lambda_s$ and $\lambda_l=2\lambda_s$. In terms of the short-lattice spacing $a_s=\lambda_s/2$, the vertical potential experienced by state $\gamma$ is
\begin{equation}
V_\gamma^z(z)=
V_{s,\gamma}\cos^2\left(\frac{\pi z}{a_s}\right)
+
V_{l,\gamma}\sin^2\left(\frac{\pi z}{2a_s}\right).
\label{eq:vertical_potential}
\end{equation}
The two components jointly determine the double-well geometry: increasing the long-period contribution  raises the barrier between neighboring bilayers while lowering the barrier within a bilayer, see App.~\ref{app:vertical_wavelength}. Because the state-dependent polarizability ratios generally differ at $\lambda_s$ and $\lambda_l$, the $g$ and $e$ states can therefore experience substantially different double-well shapes. This provides a tuning knob to engineer a strong hierarchy $t_{\perp}^{zz}=t_{\perp}^{gg}\gg t_{\perp}^{ee}=t_{\perp}^{xx}$ while simultaneously suppressing inter-bilayer leakage. The relevant quantities are thus the combined state dependence of both superlattice components, rather than the polarizability contrast at $\lambda_s$ alone. 

We choose $\lambda_s=750$~nm (and correspondingly $\lambda_l=1500$~nm) as a convenient point. As discussed in App.~\ref{app:vertical_wavelength}, this provides a favorable compromise between realizing the required intra-bilayer hopping hierarchy and isolating neighboring bilayers. At these fixed wavelengths, the $e$-state lattice depths are determined from the corresponding $g$-state depths through the polarizability ratios. The vertical sector therefore contains two independent controls, $V_{s,g}$ and $V_{l,g}$, which are adjusted to reproduce the two target ratios $t_{\perp}^{g}/t_{\parallel}^{e}$ and $t_{\perp}^{e}/t_{\parallel}^{e}$ (see Sec.~\ref{subsec:self_consistent}). Other wavelength pairs are also possible provided they generate the required state-dependent double-well geometry, discussed further in App.~\ref{app:vertical_wavelength}.

\subsection{Hybridization from a $d$-wave Rabi mask}
\label{subsec:mask}

The remaining single-particle ingredients are the inter-orbital hybridization and the crystal-field splitting,
\begin{equation}
\begin{aligned}
\hat{\mathcal{H}}_{\mathrm{hybr.}} ={}& -\sum_{\langle ij\rangle\ell\sigma} t^{xz}_{ij} \left(
\hat{c}^\dagger_{i\ell x\sigma}\hat{c}_{j\ell z\sigma} +\mathrm{h.c.} \right) \\ &+ \frac{\Delta}{2} \sum_{i\ell\sigma} \left( \hat{n}_{i\ell x\sigma} - \hat{n}_{i\ell z\sigma} \right).
\label{eq:hybr}
\end{aligned}
\end{equation}
Under the orbital assignment introduced above, the first term must convert an atom between the two clock states while simultaneously moving it to a neighboring in-plane site. While a clock drive can coherently couple the $g \leftrightarrow e$ transition, a spatially uniform field would predominantly generate \emph{on-site} conversions, which are absent in the target nickelate Hamiltonian. We therefore require a spatially structured clock field with vanishing on-site amplitude, leading nonzero matrix elements connecting nearest-neighbor sites, and signs that change between bonds along $x$ and $y$, $t^{xz}_{i,i+\hat e_x}=-t^{xz}_{i,i+\hat e_y}$. 
Engineering these three properties simultaneously is the core of the Rabi-mask construction, which we now introduce.

\begin{figure*}[t]
    \centering
    \includegraphics[width=0.95\textwidth]{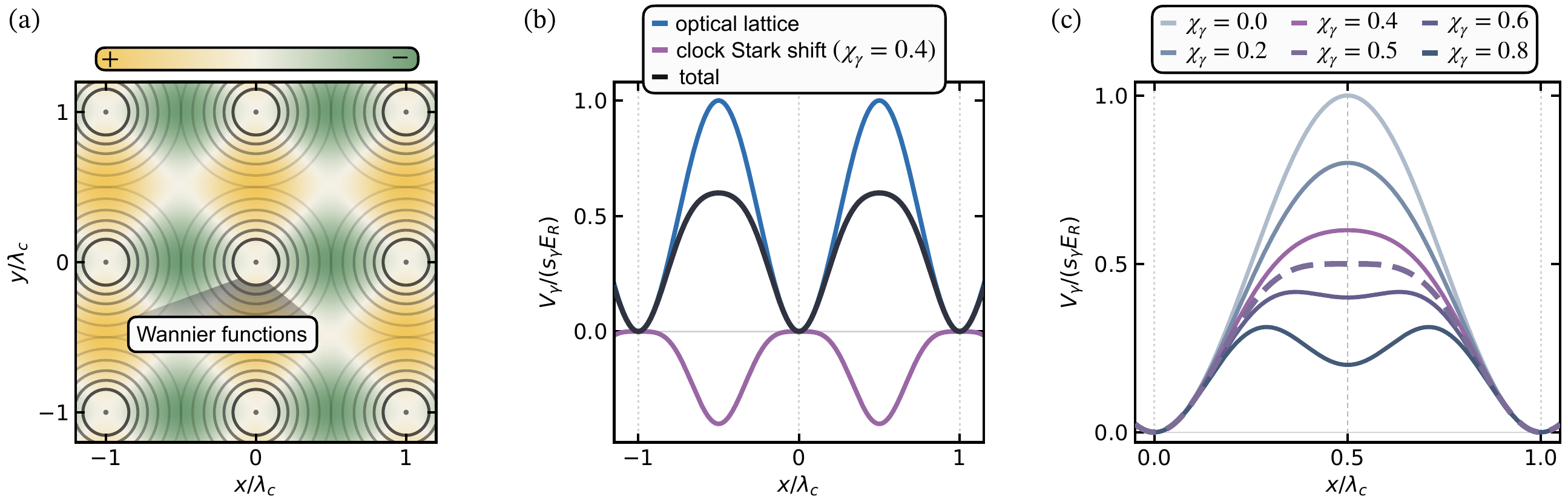}
    \caption{
    \textbf{Construction of the $d$-wave Rabi mask.}
    (a) Structured Rabi mask
    $m(\mathbf{r})
    =[\cos(Qx)-\cos(Qy)]$
    for $Q=2\pi/\lambda_c$.
    The atomic lattice is aligned such that the lattice sites coincide with nodes of the Rabi field; this way, neighboring $x$ and $y$ bonds pick up opposite Rabi signs.
    (b) Line cut along $y=0$ through the state-dependent optical-lattice potential (blue), the off-resonant Stark potential generated by the clock beams (purple), and their sum (black). The same spatial clock field that produces the coherent hybridization therefore also lowers the NN tunneling barrier through its intensity profile.
    (c) Total line-cut potential for increasing clock-light strength, parametrized by $\chi_\gamma=16v_\gamma/s_\gamma$. Increasing $\chi_\gamma$ lowers the bond-center barrier until its curvature vanishes at $\chi_\gamma=1/2$; for $\chi_\gamma>1/2$, additional bond-centered minima appear, and the single-site tight-binding description stops applying.
    }
    \label{fig:rabi_mask}
\end{figure*}

\subsubsection{Spatially structured clock coupling and lattice alignment}
Rather than driving the clock transition uniformly, we propose a two-dimensional standing-wave pattern from two retro-reflected clock-beam pairs propagating along the $x$ and $y$ directions. The beams are $\pi$ polarized with respect to the quantization axis, such that the clock drive preserves $m_F$ (which encodes the spin degree of freedom $\sigma$, discussed in Sec.~\ref{subsec:spin_encoding}). 
The clock-laser frequency must also account for the energy difference due to the state-dependent lattice. 
Choosing the optical phases of the two standing waves to differ by $\pi$ imprints a spatially structured Rabi frequency of the $g\leftrightarrow e$ transition to the atoms, 
\begin{equation}
    \Omega_\sigma(\mathbf{r}) = 2\Omega_{p,\sigma} \underbrace{[\cos(Qx) - \cos(Qy)]}_{m(\boldsymbol{r})},
\label{eq:Rabi_mask}
\end{equation}
where $\Omega_{p,\sigma}$ denotes the single-beam peak Rabi-frequency for nuclear spin state $\sigma$; the spatial structure is encoded in the mask function $m(\boldsymbol{r})$ with wavenumber $Q = {2\pi/\lamc}$. The factor of two arises from retro-reflection. The crucial feature is that the coherent Rabi coupling is linear in the optical electric-field amplitude and therefore retains its phase and sign. Under exchange of $x$ and $y$, the mask changes sign, $m(y,x)=-m(x,y)$,
and consequently transforms as the $B_{1g}$, or $d_{x^2-y^2}$, representation of the square lattice -- precisely the symmetry required in nickelate hybridization.

The real-space structure of the Rabi field is summarized in Fig.~\ref{fig:rabi_mask}~(a), showing the signed Rabi mask of Eq.~\eqref{eq:Rabi_mask}. A central aspect of our construction is to align the atomic lattice (imposed by the optical trapping lattice) with the nodal structure of the Rabi field, such that every lattice site coincides with a zero of the clock coupling. When choosing the physical in-plane lattice spacing to be $a=\lamc$, this can be achieved: With $Q=2\pi/\lamc$, it implies $Qa=2\pi$, and hence for lattice positions $\bm R_i=a(n_x,n_y)$, $m(\bm R_i)=0$. The above is illustrated in Fig.~\ref{fig:rabi_mask}~(a), where Wannier orbitals of the optical lattice are chosen to coincide with one of the two square lattices formed by the nodes of the Rabi mask with $a = \lamc$.

In contrast, at the centers of nearest neighbor (NN) bonds, $m\left(\bm R_i+\frac{a}{2}\hat e_{x/y}\right) = \mp 2$.
Thus, while the coherent clock coupling vanishes at the lattice sites, its magnitude is maximal at the centers of NN bonds. Moreover, the coupling has opposite signs on bonds oriented along $x$ and $y$. After projection onto the lattice, the spatially structured internal-state conversion therefore becomes a nearest-neighbor inter-orbital tunneling process with the required $d_{x^2-y^2}$ sign structure. In this way, the spatial phase pattern of the clock field directly encodes the orbital-symmetry structure of the nickelate hybridization.    

As mentioned in the previous subsection, to tune the in-plane hopping ratios, the in-plane lattice wavelength $\lambda_L$ is used as an active tuning parameter. While conventional counter-propagating lattices have a lattice spacing of $\lambda_L/2$, an accordion lattice can lift this constraint: by intersecting two lattice beams at a half-angle $\theta$, the physical spacing of the resulting lattice is  $a = \lambda_L/(2\sin\theta)$, where enforcing that $\theta=\arcsin[ {\lambda_L/(2\lamc)} ]$ gives $a=\lamc$. The optical wavelength and the physical lattice spacing can, therefore, be controlled independently: $\lambda_L$ selects the state-dependent polarizability ratio, while $\theta$ fixes the atomic positions relative to the clock mask. Experimentally, the relative spatial phases of the accordion lattice and the two clock standing waves must be stabilized so that the mask nodes remain aligned to the lattice sites.

With the physical spacing now fixed by the clock wavelength, it is convenient to introduce the corresponding recoil energy,
\begin{equation}
    \ER=\frac{h^2}{8ma^2} \simeq h \times 1177~\mathrm{Hz},
\label{eq:recoil}
\end{equation}
for $^{87}$Sr and $a=\lamc=698.45~\mathrm{nm}$. We use $\ER$ as the reference scale for in-plane energies below.

\subsubsection{Coherent coupling and clock-induced Stark potential.}
In a frame rotating at the laser drive frequency $\omega_d$, and within the rotating-wave approximation, the resonantly driven $g\leftrightarrow e$ transition is described by
\begin{align}
    \hat{\mathcal H}_{\rm Rabi}
    ={}&
    \frac{\hbar}{2}
    \sum_\sigma
    \int d^3r\,
    \Omega_\sigma(\mathbf{r})
    \hat\psi^\dagger_{e\sigma}(\rvec)
    \hat\psi_{g\sigma}(\rvec)
    +\mathrm{h.c.}
    \nonumber\\
    &+
    \hbar\delta
    \sum_\sigma
    \int d^3r\,
    \hat\psi^\dagger_{e\sigma}(\rvec)
    \hat\psi_{e\sigma}(\rvec),
    \label{eq:Rabi_hamiltonian}
\end{align}
where $\hat\psi_{\gamma\sigma}(\rvec)$ annihilates an atom at position $\rvec$ in state $\gamma\in\{g,e\}$ with nuclear spin $\sigma$, 
and $\delta$ is the detuning of the clock laser drive.

However, the clock beams have a second, physically distinct effect. In addition to the resonant $g\leftrightarrow e$ coupling in Eq.~\eqref{eq:Rabi_hamiltonian}, the same light couples both clock states off-resonantly to other atomic levels and thereby produces an ac Stark potential. The resonant clock transition itself is \emph{not} included in this perturbative Stark shift; its effect is already contained nonperturbatively in Eq.~\eqref{eq:Rabi_hamiltonian}. In contrast to the Rabi coupling, which is proportional to the electric-field amplitude and therefore retains the sign of the mask $m(\mathbf r)$, the ac Stark shift is proportional to the optical intensity and hence to $m^2(\mathbf r)$.

To make the relative strength of this additional potential explicit, we write
\begin{equation}
    \frac{V^{c}_\gamma(\mathbf r)}{\ER}
    =
    -4v_\gamma\,m^2(\mathbf r),
    \label{eq:clock_stark}
\end{equation}
where $\gamma\in\{g,e\}$. Here $v_\gamma$ is a dimensionless measure of the clock-light Stark shift for state $\gamma$, expressed in units of the in-plane recoil energy $\ER$. More precisely, for a single-beam peak clock intensity $I_p$,
\begin{equation}
    v_\gamma
    =
    \frac{\kappa_\gamma I_p}{\ER/h},
    \qquad
    \kappa_\gamma
    =
    \frac{\alpha^{\mathrm{off}}_\gamma(\lambda_c)}
    {2\epsilon_0 c h},
    \label{eq:v_gamma}
\end{equation}
where $\alpha^{\mathrm{off}}_\gamma(\lambda_c)$ denotes the off-resonant, or background, polarizability at the clock wavelength, with the explicitly driven $g\leftrightarrow e$ transition excluded. Thus $v_\gamma$ sets the overall strength of the clock-induced deformation of the lattice, while $m^2(\mathbf r)$ fixes its spatial profile.

Along $x$, the ordinary in-plane optical lattice and the clock-induced Stark contribution combine to give 
\begin{equation} 
\frac{V_\gamma(x,0)}{\ER} = s_\gamma \sin^2\left(\frac{\pi x}{a}\right) - 16v_\gamma \sin^4\left(\frac{\pi x}{a}\right), 
\label{eq:clock_stark_linecut} 
\end{equation} 
where $s_\gamma$ is the state-dependent lattice depth of the optical lattice. As shown in Fig.~\ref{fig:rabi_mask}~(b), the clock light is most attractive near the NN bond center and therefore lowers the tunneling barrier between adjacent sites. It is useful to characterize this deformation by
$\chi_\gamma = 16v_\gamma/s_\gamma$. For small $\chi_\gamma$, the clock field lowers the original barrier while leaving one site-centered minimum per unit cell. As the clock intensity is increased, however, the curvature at the bond center decreases, as shown in Fig.~\ref{fig:rabi_mask}~(c). The bond center becomes locally flat at $\chi_\gamma=1/2$, and turns into an additional local minimum for $\chi_\gamma>1/2$. This provides an upper bound on the usable clock intensity: the Rabi mask cannot be made arbitrarily strong without changing the low-energy lattice structure itself. The self-consistent treatment of this deformation is developed in Sec.~\ref{subsec:self_consistent}.

Note that, for the orbital assignment used here, the clock-induced modification of the lattice acts in the same direction as the desired in-plane hopping hierarchy. The $e$ state experiences a stronger clock-light Stark potential than the $g$ state, while at the same time occupying the shallower bare in-plane lattice. Consequently, the dimensionless deformation $\chi_\gamma=16v_\gamma/s_\gamma$ is substantially larger for $e$ than for $g$. The clock beams therefore enhance $t^{ee}_{\parallel}$ relative to $t^{gg}_{\parallel}$, reinforcing the target hierarchy $t^{xx}_{\parallel}>t^{zz}_{\parallel}$.

\subsubsection{Projection onto the lattice}
We now project the coherent part of Eq.~\eqref{eq:Rabi_hamiltonian} onto the lowest-band (in-plane) Wannier orbitals $w_{\gamma}(\mathbf r-\mathbf R_i)$ centered around lattice site $\mathbf R_i$. Expanding the field operators as
\begin{equation}
\hat{\psi}_\gamma^{(\dagger)}(\mathbf r)
=
\sum_i
w_\gamma^{(*)}(\mathbf r-\mathbf R_i)\,
\hat c^{(\dagger)}_{i\gamma},
\label{eq:wannier_expansion}
\end{equation}
the relevant overlap between a $g$ orbital centered at the origin and an $e$ orbital displaced by $\bm d$ is then
\begin{equation}
K(\bm d)
=
\int d^2 r\,
w_e^*(\mathbf{r}-\bm d)\,
m(\mathbf{r})\,
w_g(\mathbf{r}).
\label{eq:K_d}
\end{equation}
The dimensionless quantity $K(\bm d)$ contains the full spatial information required to determine the clock-assisted tunneling term.

The desired selection rules as imposed by the material now indeed follow from the symmetry of the Rabi mask. Although the coherent mask changes sign under a $C_4$ rotation, $m(C_4\mathbf{r})=-m(\mathbf{r})$, the total trapping potential, including the clock-induced Stark contribution of Eq.~\eqref{eq:clock_stark}, remains $C_4$ symmetric. Consequently, $K(C_4\bm d)=-K(\bm d)$. Likewise, reflection across the diagonal, $\mathcal M_{xy}:(x,y)\mapsto(y,x)$, gives $K(\mathcal M_{xy}\bm d)=-K(\bm d)$. Applying these relations to the relevant lattice vectors yields
\begin{equation}
\begin{gathered}
K(\bm 0)=0,\\
K(a\hat e_y)=-K(a\hat e_x),\\
K(a\hat e_x+a\hat e_y)=0.
\label{eq:selection_rules}
\end{gathered}
\end{equation}
The spatial symmetry of the material orbitals is thus directly transferred onto the synthetic orbital dimensions of the simulated Hamiltonian, without requiring Wannier orbitals of the same microscopic shape.

The leading nonzero matrix element is therefore the NN bond overlap $K_1\equiv K(a\hat e_x)$, which gives
\begin{equation}
t^{xz}_{i, i+\hat{e}_x; \sigma}
=
+\hbar\Omega_{p,\sigma} K_1,
\qquad
t^{xz}_{i, i+ \hat{e}_y; \sigma}
=
-\hbar\Omega_{p,\sigma} K_1.
\label{eq:spin_hybridization}
\end{equation}
The overall sign can be reversed through the global optical phase, while the relative minus sign between $x$ and $y$ bonds is fixed by the symmetry of the mask. Fourier transforming the NN term therefore gives an inter-orbital coupling proportional to $\cos(k_xa)-\cos(k_ya)$, exactly reproducing the $B_{1g}$ form factor of the target Hamiltonian.

\subsubsection{Crystal-field control from clock detuning}
The detuning term in Eq.~\eqref{eq:Rabi_hamiltonian} provides an independent control of the relative orbital energy. Since the $e$ state encodes the $x=\dxy$ orbital and the $g$ state encodes the $z=\dz$ orbital,
\begin{equation}
\hbar\delta\,\hat n_e
=
\frac{\hbar\delta}{2}
\left(
\hat n_e-\hat n_g
\right)
+
\frac{\hbar\delta}{2}
\left(
\hat n_e+\hat n_g
\right).
\end{equation}
The second term is proportional to the conserved total density and can be absorbed into the chemical potential. Comparing the first term with Eq.~\eqref{eq:hybr} therefore gives
\begin{equation}
\Delta \:\:\leftrightarrow\:\: \hbar\delta.
\label{eq:crystal_field_detuning}
\end{equation}
The hybridization amplitude and crystal field are thus controlled by two independent properties of the clock drive: its intensity determines $\Omega_p$ and hence $t^{xz}$, while its frequency determines $\delta$ and hence $\Delta$.

\subsection{Interaction structure}
\label{sec:int_AEA}

We now turn to the interaction sector of the simulator. For AEA at the temperatures of interest, interactions are dominated by short-range $s$-wave collisions. Projecting the corresponding contact interaction onto the localized three-dimensional Wannier orbitals $W_\gamma = w_{\gamma}(x,y) \phi_\gamma(z)$ therefore generates on-site interactions of the form
\begin{equation}
    U_{\gamma\gamma'}
    =
    \frac{4\pi\hbar^2 a_{\gamma\gamma'}}{m}
    \int d^3r\,
    |W_\gamma(\rvec)|^2
    |W_{\gamma'}(\rvec)|^2,
    \label{eq:contact_projection}
\end{equation}
where $\gamma,\gamma'\in\{g,e\}$ label the two clock states, $a_{\gamma\gamma'}$ denotes the corresponding $s$-wave scattering length, and $\phi_\gamma(z)$ is the Wannier orbital of a single isolated bilayer of the vertical lattice. 

Collisions between two atoms in the same electronic state generate the intra-orbital Hubbard interactions $U_{gg}$ and $U_{ee}$. For one atom in $g$ and one in $e$, there are instead two scattering channels with interaction energies $U_{eg}^{+}$ and $U_{eg}^{-}$, corresponding to symmetric and antisymmetric combinations of the electronic states under particle exchange. This leads to a direct inter-orbital interaction and an orbital-exchange interaction~\cite{gorshkov2010},
\begin{equation}
    V
    =
    \frac{U_{eg}^{+}+U_{eg}^{-}}{2},
    \qquad
    \Jex
    =
    \frac{U_{eg}^{+}-U_{eg}^{-}}{2}.
    \label{eq:V_Jex_definition}
\end{equation}
In $\Sr$, the ratio of exchange to direct interactions is natively set by the background atomic scattering lengths
\begin{equation}
    \frac{\Jex}{V}
    =
    \frac{
        a_{eg}^{+}-a_{eg}^{-}
    }{
        a_{eg}^{+}+a_{eg}^{-}
    },
    \label{eq:J_over_V}
\end{equation}
that already achieves the interaction hierarchy of Eq.~(\ref{eq:target_interactions}), see Sec.~\ref{subsec:self_consistent}. The overall interaction scale relative to the tunneling can be varied through the lattice depths, since stronger confinement increases the Wannier-overlap integrals in Eq.~\eqref{eq:contact_projection}. 
Meanwhile, tuning the ratio of ${\Jex / V}$ might be achieved through optical Feshbach resonances \cite{Theis04_PRL, Blatt11_PRL}, although we leave further discussions of such control elsewhere. 

\begin{figure*}
\includegraphics[width=0.9\textwidth]{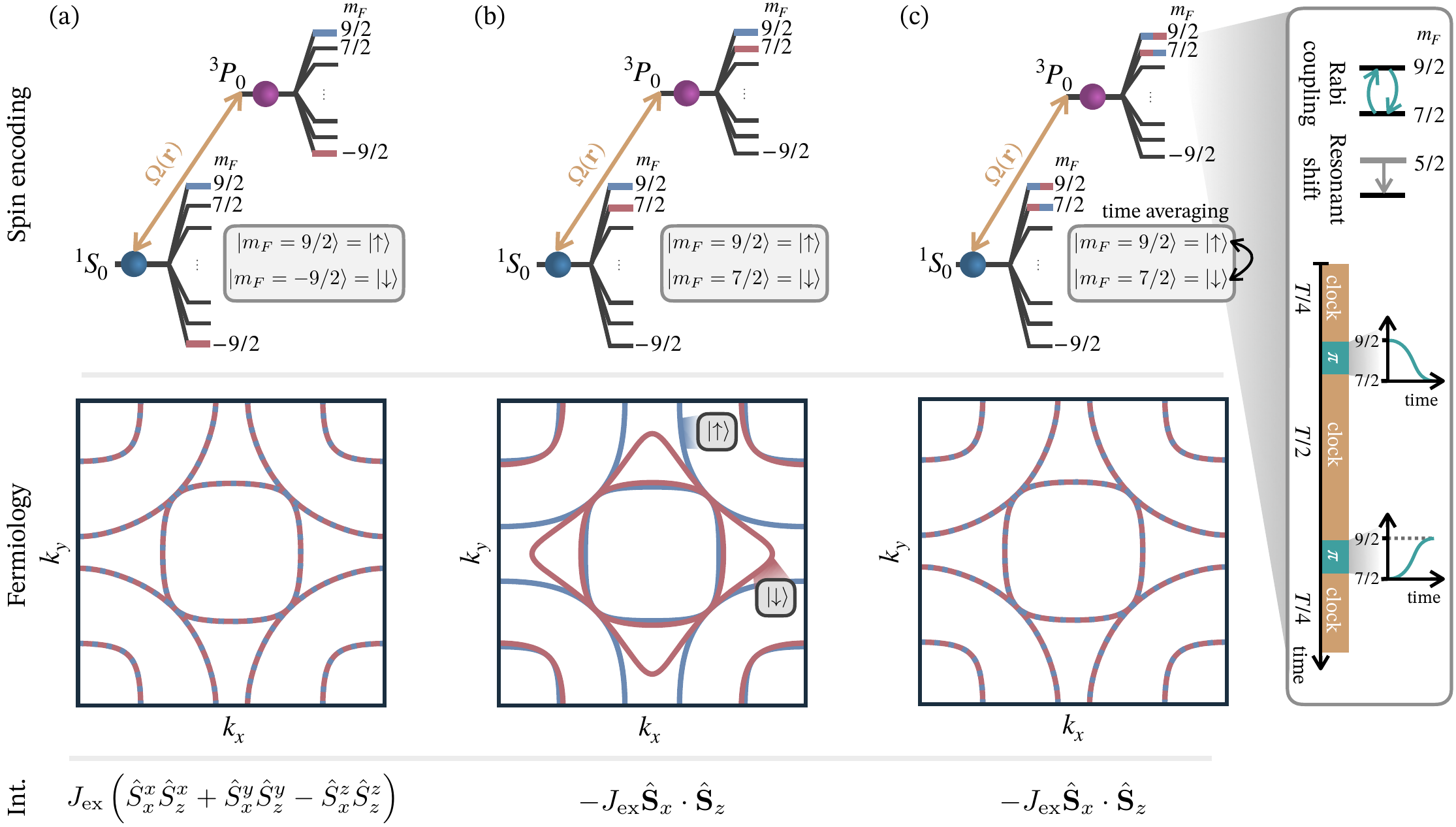}
\caption{
\textbf{Nuclear-spin encoding as a knob for engineering Hund coupling.}
(a) Encoding $\uparrow,\downarrow$ in the pair $m_F=\pm9/2$ gives equal hybridization magnitudes with opposite signs. Gauging the hybridization to a common sign leaves the fermiology unchanged but transforms the isotropic Hund interaction into an anisotropic exchange. (b) Choosing the adjacent states $m_F=9/2,7/2$ preserves the native isotropic ferromagnetic Hund coupling, at the cost of unequal clock matrix elements and hence spin-dependent hybridization amplitudes. (c) This mismatch can be removed by periodically exchanging the two nuclear-spin states during the clock evolution. The resulting time-averaged Hamiltonian combines equal effective hybridization for both spins with the native isotropic Hund interaction; the pulse sequence is shown schematically on the right.
}
\label{fig:spinenc}
\end{figure*}

After projection onto the lowest lattice orbitals, the resulting collisional Hamiltonian is
\begin{equation}
\begin{aligned}
    \hat{\mathcal H}_{\mathrm{coll}}
    ={}&
    \sum_{i\ell}
    \left[
        U_{gg}\,
        \hat n_{i\ell g\uparrow}
        \hat n_{i\ell g\downarrow}
        +
        U_{ee}\,
        \hat n_{i\ell e\uparrow}
        \hat n_{i\ell e\downarrow}
    \right]
    \\
    &+
    V
    \sum_{i\ell\sigma\sigma'}
    \hat n_{i\ell e\sigma}
    \hat n_{i\ell g\sigma'}
    \\
    &+
    \Jex
    \sum_{i\ell\sigma\sigma'}
    \hat c^\dagger_{i\ell e\sigma}
    \hat c^\dagger_{i\ell g\sigma'}
    \hat c_{i\ell e\sigma'}
    \hat c_{i\ell g\sigma}.
    \label{eq:collisional_hamiltonian}
\end{aligned}
\end{equation}
Because the electronic and nuclear-spin degrees of freedom are nearly decoupled in AEA, the scattering parameters are to a very good approximation independent of $\sigma$. The interaction therefore naturally inherits the spin symmetry required for the two-orbital model.

For $\Sr$, the relevant scattering lengths are
$a_{gg}=96.2a_0, a_{ee}=176a_0, a_{eg}^{+}=169a_0, a_{eg}^{-}=68a_0$, with $a_0$ the Bohr radius~\cite{zhang2014,goban2018,Bilokon2023}. Since $a_{eg}^{+}>a_{eg}^{-}$, Eq.~\eqref{eq:V_Jex_definition} yields $\Jex>0$; a ferromagnetic Hund exchange is therefore generated directly by the atomic collision physics, without requiring additional engineering.

The simulator thus reproduces the essential terms of local two-orbital interactions: intra-orbital Hubbard repulsions, a direct inter-orbital repulsion, and a ferromagnetic Hund exchange. Compared to Eq.~\eqref{eq:Hint}, only the pair-hopping term of the full Hubbard-Kanamori Hamiltonian is not generated by the contact interaction in the present implementation; this, however, is not expected to alter the qualitative pairing physics of the bilayer nickelate model at the relevant interaction strengths, where doubly occupied sites relevant for the pair hopping occur only virtually~\cite{WangZhang2025}. 

\subsection{Nuclear-spin encoding}
\label{subsec:spin_encoding}

The final degree of freedom that must be encoded into the simulator is electron spins. 
We choose to do so with two magnetic sub-levels of the $F=9/2$ hyperfine manifold. Because the alkaline-earth interactions are SU($N$) symmetric to high accuracy, none of the collisional parameters of Sec.~\ref{sec:int_AEA} depend on the selected pair. The choice does, however, enter the Rabi frequency driving the clock transition
\begin{align}
    \Omega_{p, m_F} 
    =
    \eta_{m_F} \Omega_0, 
\end{align}
since the $\pi$-polarized coupling strength depends on $m_F$ through the Clebsch-Gordan coefficient
\begin{equation}
    \eta_{m_F} 
    =
    \frac{m_F}{\sqrt{F(F+1)}},
 \label{eq:eta_mf}
\end{equation}
where the Rabi amplitude is given by the product of the reduced transition dipole matrix element and the laser field strength $\Omega_0 = d_{g e} E_{c} / \hbar$. The spatial form factor $m(\boldsymbol{r})$ defined in Eq.~\eqref{eq:Rabi_mask} is common to all spin states, but the sign and magnitude of the hybridization inherit $\eta_{m_F}$. This dependence on $m_F$ translates to a spin-dependent inter-orbital hopping term, providing a handle to deform the Hund and hybridization sectors independently. 

For our current focus on La$_3$Ni$_2$O$_7$, engineering a model that accurately emulates this material leads to two natural spin encodings that we now describe, summarized in Fig.~\ref{fig:spinenc}.

\subsubsection{Time-reversal pair \texorpdfstring{$m_F=\pm9/2$}{mF = +/- 9/2}}

The first encoding involves 
\begin{equation}
 \up\equiv (m_F=+9/2),
 \qquad
 \dn\equiv (m_F=-9/2)
\end{equation}
which gives both spin components equal inter-orbital hopping amplitudes but with opposite signs. Suppressing the layer and bond labels, this encoding results in the hybridization term
\begin{equation}
 \hat{\mathcal{H}}_{\text{hybr.}}\sim t^{xz}
 \left(
 \hat c_x^{\dagger}\sigma_z\hat c_z
 +\mathrm{h.c.}
 \right),
 \label{eq:opposite_sign_hyb}
\end{equation}
where $\hat{c}_\gamma = (\hat{c}_{\gamma,\uparrow}, \hat{c}_{\gamma, \downarrow})$ and $\sigma_z$ is the $z$-Pauli matrix that acts in the encoded spin space. The relative sign can be removed from the single-particle Hamiltonian by the spin-dependent orbital gauge transformation
\begin{equation}
     \hat c_{z\dn}\rightarrow-\hat c_{z\dn},
     \qquad
     \hat c_{z\up}\rightarrow\hat c_{z\up},
     \label{eq:gauge_transform}
\end{equation}
that leaves all density operators unchanged. 

However, this gauge transformation does not preserve the spin-rotationally invariant form of the exchange term, transforming the exchange operator~\footnote{The transformation $\hat{c}_{z\dn}\to-\hat{c}_{z\dn}$ is a rotation by $\pi$ about the $z$ axis acting only on the $z$-orbital pseudospin, and sends $S_z^x\to-S_z^x, \quad S_z^y\to-S_z^y, \quad S_z^z\to S_z^z$. Therefore
 $-2\Jex\hat{\mathbf{S}}_x\cdot\hat{\mathbf{S}}_z
 \to
 2\Jex(S_x^xS_z^x+S_x^yS_z^y-S_x^zS_z^z)$.}
\begin{equation}
 -2\Jex\hat{\mathbf{S}}_x\cdot\hat{\mathbf{S}}_z
 \to
 2\Jex(S_x^xS_z^x+S_x^yS_z^y-S_x^zS_z^z),
\end{equation}
i.e., the transverse spin-flip terms change sign while the longitudinal term remains invariant. 

This encoding therefore provides a direct way to probe how the pairing physics responds to a modified Hund sector. In particular, several strong-coupling scenarios for bilayer nickelates rely on antiferromagnetic inter-orbital exchange as an important ingredient for pairing~\cite{lu2024, Qu2024, Oh2023, Schloemer2024_tJ}. The gauge-transformed interaction above offers a controlled route to alter the sign of the transverse Hund exchange while leaving the single-particle fermiology unchanged, making it possible to experimentally test how sensitive the pairing tendencies are to the structure and sign of the Hund coupling.

\subsubsection{Adjacent pair \texorpdfstring{$m_F=9/2,\,7/2$}{mF = 9/2, 7/2}}

A second choice is the encoding
\begin{equation}
 \up\equiv (m_F=9/2),
 \qquad
 \dn\equiv (m_F=7/2).
\end{equation}
Both clock matrix elements now have the same sign, and the native isotropic Hund interaction of Eq.~\eqref{eq:Hint} is preserved exactly. 
Instead, the tunneling magnitude is now spin-dependent,
\begin{equation}
    \frac{t^{xz}_{\up}}{t^{xz}_{\dn}}=\frac{9}{7}.
    \label{eq:nine_seven_ratio}
\end{equation}
Pinning the arithmetic mean of the two hybridizations to the material target, the two amplitudes are $t^{xz}_{\up}=\frac{9}{8}\avg{t^{xz}}$, $
 t^{xz}_{\dn}=\frac{7}{8}\avg{t^{xz}}$,
with $\avg{t^{xz}} = t^{xz}_\parallel$. This asymmetry is not removable by a gauge transformation; it produces spin-dependent band dispersions and, at fixed total density, a Fermi-surface mismatch between the two encoded spins, see Fig.~\ref{fig:spinenc}~(b). 

The two encodings above thus feature different deformations of the target model. The $\pm9/2$ pair preserves equal band magnitudes and the time-reversal pairing of the two spin species, but changes the sign structure of the Hund coupling once the hybridization is gauged to a common sign. The $9/2,7/2$ pair preserves the native Hund interactions and a common hybridization sign, but explicitly breaks spin symmetry in the band Hamiltonian. Because the two deformations act in different sectors, comparing them provides an experimental test of which features of the target model control its low-temperature physics. 

\subsubsection{Floquet-engineered pair \texorpdfstring{$m_F=9/2,\,7/2$}{mF = 9/2, 7/2} }

Finally, we propose a pulse-engineered encoding that fixes the amplitude mismatch in the $m_F=9/2, 7/2$ encoding to effectively achieve $t^{xz}_{\up} = t^{xz}_{\dn}$. We propose to use pulses instead of continuous laser drives in order to minimize heating from photon scattering and spontaneous emission. A simple pulse protocol involves the symmetric sequence illustrated in the inset of Fig.~\ref{fig:spinenc}~(c), where three segments of coherently driving the clock transition (labeled ``clock") are interspersed by rapid $\pi$-pulses that flip $\up$ to $\dn$ and vice versa. The duration of each clock segment is determined by the total time of the Floquet pulse sequence $T$. The resulting effective Floquet Hamiltonian, obtained from average Hamiltonian theory~\cite{Magnus1954, BlanesRos2009}, yields
\begin{equation}
    t_{\up}^{xz,\mathrm{eff}} \approx t_{\dn}^{xz,\mathrm{eff}} \approx \frac{t_{\up}^{xz}+t_{\dn}^{xz}}{2}.
\end{equation}

Due to the small nuclear magnetic moment, these rapid $\pi$-pulses are most practically implemented with a two-photon Raman transition, passing through a far-detuned but dipole-coupled intermediate state. These $\pi$-pulses must be implemented on each clock state separately, with the spin states in $^{1}S_0$ Raman coupled to $|^{3}P_1; F=9/2, m_F=7/2\rangle$, while those in $^{3}P_0$ Raman coupled to $|^{3}D_1; F=9/2, m_F=7/2\rangle$.   

To suppress leakages to unwanted $m_F$ states, we are also required to detune the other magnetic sub-levels away to isolate the $m_F=9/2$ and $7/2$ states in each clock manifold. This detuning can be achieved through a state-selective resonant drive on $m_F = 5/2$, which induces the required Autler-Townes splitting as was demonstrated on the $m_F=-5/2$ state in the $^{1}S_0$ manifold~\cite{Barnes2022}. Further details and a theoretical analysis of this scheme are provided in App.~\ref{app:pulse_engineered_tunneling}, showing that an effective Floquet Hamiltonian with  $t^{xz, \mathrm{eff}}_{\up} = t^{xz, \mathrm{eff}}_{\dn}$ can be obtained with currently accessible laser generation technologies.  

\subsection{Self-consistent Wannierization}
\label{subsec:self_consistent}

Because the clock field both generates the inter-orbital hybridization and modifies the optical potential through its off-resonant Stark shift, the Wannier orbitals entering the hybridization matrix element depend on the clock-drive strength. The hopping and hybridization targets must therefore be imposed self-consistently. For state $\gamma\in\{g,e\}$, the full in-plane potential is
\begin{align}
\frac{V_\gamma(x,y)}{\ER}={}&
s_\gamma
\left[
\sin^2\left(\frac{\pi x}{a}\right)
+
\sin^2\left(\frac{\pi y}{a}\right)
\right]
\nonumber\\
&-
4v_\gamma
\left[
\cos\left(\frac{2\pi x}{a}\right)
-
\cos\left(\frac{2\pi y}{a}\right)
\right]^2 .
\label{eq:full_inplane_potential}
\end{align}

We use the single-beam peak Rabi frequency $\Omega_p$ as the clock-drive parameter. The associated dimensionless Stark strength is
\begin{equation}
v_\gamma(\Omega_p)
=
\frac{\alpha_\gamma^{\rm off}(\lambda_c)}{4\ER}
\left(
\frac{\hbar\Omega_p}{d_{eg}}
\right)^2 ,
\end{equation}
where $\alpha_\gamma^{\rm off}(\lambda_c)$ is the background polarizability at the clock wavelength with the resonantly driven $g\leftrightarrow e$ transition excluded.

For a trial value of $\Omega_p$, the calculation proceeds as follows: First, the lowest Bloch band of Eq.~\eqref{eq:full_inplane_potential} is obtained in a two-dimensional plane wave basis. The hopping amplitudes are then extracted from the band dispersion,
\begin{equation}
t_{\gamma,\bm d}
=
-\frac{1}{N_k}
\sum_{\kvec}
E_{0\gamma}(\kvec)e^{-i\kvec\cdot\bm d}.
\label{eq:hopping_fourier}
\end{equation}
The phase of each Bloch state is fixed such that its amplitude at the site center is real and positive, and the corresponding Bloch sum is used to construct a site-centered Wannier function. Finally, the updated Wannier functions are used to evaluate the nearest-neighbor clock overlap $K_1$.

From $K_1$, the hybridization strength $|t^{xz}_{\parallel}| = \hbar\Omega_p |K_1|$ can be calculated. With $r_{xz}=0.495$ denoting the target ratio $t^{xz}_{\parallel}/t^{xx}_{\parallel}$ from Table~\ref{tab:hoppings}, the Rabi frequency required by the Wannier functions obtained at a trial value $\Omega_p$ is therefore
\begin{equation}
\Omega_{p,\mathrm{req}}(\Omega_p)
=
\frac{
r_{xz}}{|K_1(\Omega_p)|}\frac{t^{xx}_{\parallel}(\Omega_p)
}{\hbar }.
\label{eq:Omega_required}
\end{equation}
The target value satisfies $\Omega_{p,\mathrm{req}}(\Omega_p)=\Omega_p$ and is obtained iteratively. Each evaluation requires complete two-dimensional Bloch and Wannier calculations for both clock states.

The clock loop is nested inside the hopping optimization. At fixed
in-plane depth $s_e=s_x$, the lattice wavelength $\lambda_L$ is varied
until $t^{zz}_{\parallel}/t^{xx}_{\parallel}$ reaches its target, with
$\Omega_p$ re-converged at every trial wavelength. The two vertical
depths $V_{s,g}$ and $V_{l,g}$ are then adjusted to reproduce the two
interlayer hopping ratios at fixed $\lambda_s = 750$~nm. Finally,
the interaction integrals are evaluated using the converged
in-plane Wannier functions $w_\gamma(x,y)$ together with the localized single-well orbitals $\phi_\gamma(z)$ of the vertical dimer. More details of the self-consistent Wannierization are presented in App.~\ref{app:wannier}.

\begin{figure}[t]
\centering
\includegraphics[width=\columnwidth]{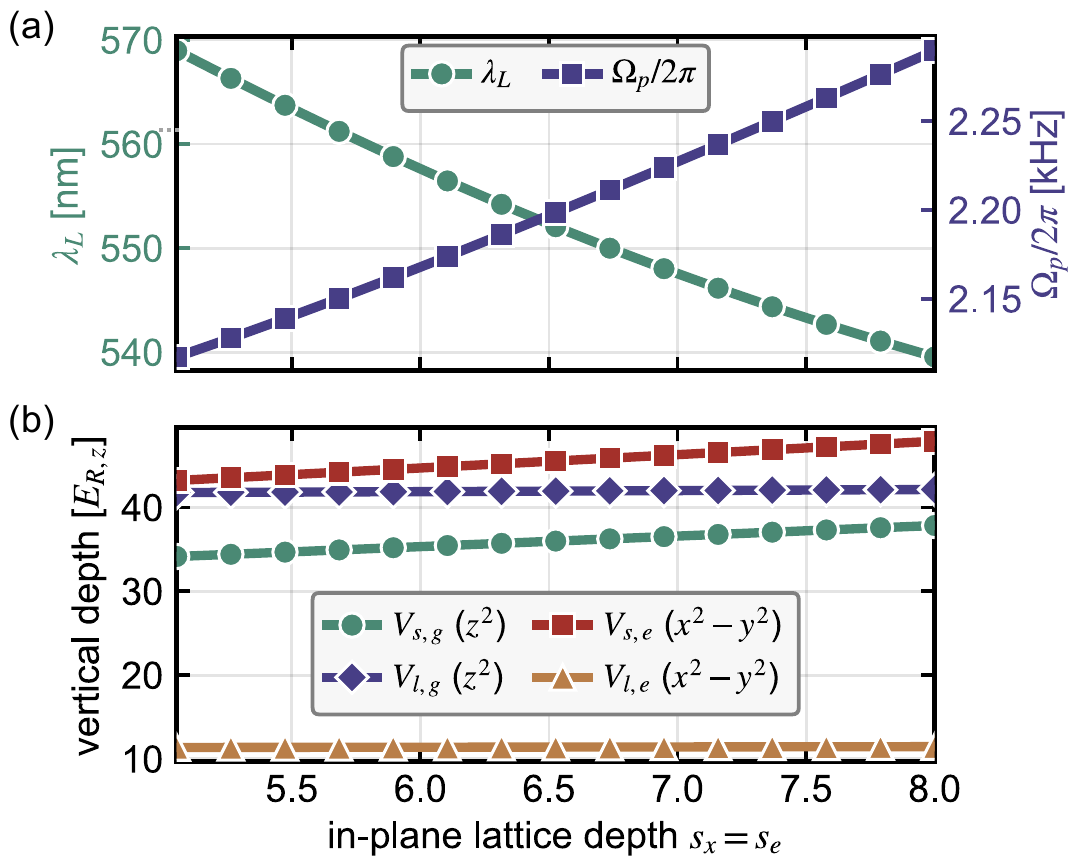}
\caption{\textbf{Optimized experimental parameters.}
(a) Optimized in-plane trapping wavelength $\lambda_L$ and single-beam clock Rabi frequency $\Omega_p/2\pi$ as a function of the in-plane lattice depth $s_x=s_e$. (b) Corresponding optimized depths of the short- and long-wavelength components of the vertical state-dependent superlattice. The two lattice depths are adjusted to reproduce the target interlayer hopping amplitudes for the $d_{z^2}$ and $d_{x^2-y^2}$ orbitals while maintaining isolated bilayers.}
\label{fig:opt_params}
\end{figure}

Figure~\ref{fig:optimization_summary} summarizes the main result of the optimization, as also discussed in Sec.~\ref{sec:summary}. In Fig.~\ref{fig:optimization_summary}~(a), we demonstrate that the full target hopping structure from Tab.~\ref{tab:hoppings} (dotted gray lines) can be realized over a wide range of in-plane lattice depths. For each value of $s_x=s_e$, the remaining experimental parameters are readjusted self-consistently such that all hopping ratios, including the clock-induced hybridization $t^{xz}_{\parallel}$, remain fixed at their target values. Over the displayed range, the dominant in-plane hopping is
$t^{ee}_{\parallel}/h = t^{xx}_{\parallel}/h \simeq 50$-$100$~Hz.

Varying the in-plane lattice depth while keeping the hopping Hamiltonian fixed provides a direct handle on the interaction regime [Fig.~\ref{fig:optimization_summary}(b)]. For the shallower lattices, the intra-orbital interactions $U_x$ and $U_z$ are nearly identical, but become more separated at larger depths. Across the full displayed range, the desired hierarchy between $U_x$, $U_z$, $V$, and $J_{\mathrm{ex}}$ remains unchanged, while their magnitudes relative to $t^{xx}_{\parallel}$ vary substantially. The same hopping Hamiltonian can therefore be explored across qualitatively different interaction strengths without changing its underlying band structure. 

The corresponding optimized in-plane wavelengths and clock Rabi frequencies are plotted in Fig.~\ref{fig:opt_params}(a). 
Similar Rabi frequencies of the required magnitude have been demonstrated in Ref.~\cite{Yan25_PRX}. The vertical lattice depths required to maintain the target intra-bilayer hoppings are shown in Fig.~\ref{fig:opt_params}(b), in units of the vertical recoil energy $E_{R,z}$ for $\lambda_s=750$~nm ($\lambda_l=1500$~nm). The required depths remain in a standard experimentally accessible range throughout the scan. As we show in App.~\ref{app:vertical_wavelength}, inter-bilayer leakage is indeed minimized at the here chosen vertical lattice wavelength $\lambda_s = 750$~nm, with sub-Hz tunneling rates between isolated bilayers. 

The lower end of the in-plane lattice depth scan in Figs.~\ref{fig:optimization_summary} and~\ref{fig:opt_params} is set by the $\chi_\gamma < 1/2$ condition. Although the required clock drive decreases toward shallower lattices, the bare lattice confinement decreases more rapidly, so that the relative deformation $\chi_\gamma$ increases (primarily of the excited clock state, $\chi_e$). Below the lattice depths shown in Figs.~\ref{fig:optimization_summary} and~\ref{fig:opt_params}, $\chi_e\geq 1/2$, causing the clock-induced potential to develop an additional bond-centered minimum [Fig.~\ref{fig:rabi_mask}(c)].

\section{Conclusions}
\label{sec:conclusions}
We present an architecture for the quantum simulation of material-realistic multi-orbital Hamiltonians with ultracold alkaline-earth atoms. By combining long-lived clock states as synthetic orbitals, state-dependent optical potentials, spatially structured coherent coupling, and ultracold collisional interactions, the proposed platform provides control over orbital-selective tunneling, multi-orbital interactions, symmetry-selective hybridization, and crystal-field splittings.

We showcase these capabilities by mapping the Hamiltonian of bilayer nickelate La$_3$Ni$_2$O$_7$ onto ultracold $\Sr$ atoms. The resulting simulator reproduces the three key features of the target single-particle Hamiltonian: a strongly dispersive $\dxy$ orbital in the plane, a $\dz$ orbital with dominant tunneling across the bilayer, and the nearest-neighbor $\Bog$ inter-orbital hybridization. 
The correspondence between the target material Hamiltonian and the experimentally controlled simulator parameters is summarized in Tab.~\ref{tab:dictionary}. 
We envision these quantum simulation experiments to be performed under a quantum gas microscope, allowing site- and orbital-resolved measurements for mapping out the phase diagram.

The target hopping ratios, including the clock-laser-induced inter-orbital hybridization, can be maintained over a continuous range of in-plane lattice depths. The same variation of lattice depth changes the interaction strengths substantially while preserving the desired interaction hierarchy. The hopping structure can therefore be held fixed while the system is moved through varying correlation regimes.

\begin{table}[t]
\caption{Comparison between the target model of Sec.~\ref{sec:model} and the alkaline-earth implementation.}
\label{tab:dictionary}
\begin{ruledtabular}
\begin{tabular}{lll}
target & simulator & control \\
\hline
$x=\dxy$ & $e={}^3P_0$ & fixed by Eq.~\eqref{eq:swapped_mapping} \\
$z=\dz$ & $g={}^1S_0$ & fixed by Eq.~\eqref{eq:swapped_mapping} \\
$\sigma=\up,\dn$ & $m_F$ pair & state preparation \\
layer $\ell$ & dimer well & vertical superlattice \\
\hline
$t^{xx}_{\parallel}$ & $t^{ee}_{\parallel}$ & in-plane depth $s_e$ \\
$t^{zz}_{\parallel}$ & $t^{gg}_{\parallel}$ & $\alpha_e/\alpha_g$ at $\lambda_L$ \\
$t^{zz}_{\perp}$ & $t^{gg}_{\perp}$ & $V_{s,g},V_{l,g}$ \\
$t^{xx}_{\perp}$ & $t^{ee}_{\perp}$ & set by $\alpha$ ratios \\
$t^{xz}_{\parallel}$ & $\hbar\Omega_p K_1$ & clock intensity $I_p$ \\
$\Delta$ & $\hbar\delta$ & clock detuning \\
\hline
$U_x$ & $U_{ee}$ & $a_{ee}$, $\int|W_e|^4$ \\
$U_z$ & $U_{gg}$ & $a_{gg}$, $\int|W_g|^4$ \\
$V$ & $(U^{+}_{eg}+U^{-}_{eg})/2$ & $a^{\pm}_{eg}$, $\int|W_e|^2|W_g|^2$ \\
$J_H$ & $(U^{+}_{eg}-U^{-}_{eg})/2$ & $a^{\pm}_{eg}$, $\int|W_e|^2|W_g|^2$ \\
\end{tabular}
\end{ruledtabular}
\end{table}

The large $I = 9/2$ nuclear spin of $\Sr$ allows for a wide selection of spin encodings to realize exotic inter-orbital hybridization. Among these options, we showed that spin-independent hybridization can be achieved by Floquet engineering, yielding the bilayer nickelate Hamiltonian with high accuracy. Other spin-dependent realizations offer complementary ways of probing the robustness of magnetic and pairing phenomena in nickelate superconductors.

More broadly, the ingredients presented here form a general recipe for engineering multi-orbital Hamiltonians. In particular, the structured clock drive provides a direct route to momentum-dependent hybridization form factors: by combining different spatial Fourier components of the optical field, one can imprint couplings transforming according to different point-group representations. Additionally, interactions can be tuned from the native interatomic scattering lengths by means of optical Feshbach resonances, to realize a wide range of interaction hierarchies. 

Our toolbox provides a particularly flexible framework for moving beyond paradigmatic single-band Hubbard models, toward material-realistic quantum simulation.  While realizing the full bilayer-nickelate Hamiltonian requires a comparatively demanding combination of state-dependent lattices, structured clock driving, and bilayer engineering, all the required ingredients are compatible with current-day tabletop laser technologies.

\begin{acknowledgments}
We thank William Milner, Simon Karch, Yann Kiefer, Kaden Hazzard, Hannah Lange, Lukas Homeier, Fabian Grusdt, Annabelle Bohrdt, Norman Yao, Chris Laumann and Ana Maria Rey for helpful discussions. We acknowledge 54TRB for discussions that helped seed the ideas developed in this work. H.S., R.R.W.W. and H.R.S. acknowledge support from the NSF through a grant for ITAMP at Harvard University. B.X. acknowledges support from the A*STAR International Fellowship. SFY would like to acknowledge the NSF via PHY-2608210 and via the CUA PFC grant (PHY-2317134) as well as the Q-SEnSE QLCI (OMA-201624).
\end{acknowledgments}

\appendix

\section{Numerical Wannier construction}
\label{app:wannier}

\subsection{Two-dimensional plane-wave Hamiltonian}
Coordinates are measured in units of the lattice spacing $a$. The plane-wave basis is
\begin{equation}
 \langle\rvec|m,n;\kvec\rangle
 =e^{i[(k_x+2\pi m)x+(k_y+2\pi n)y]},
\end{equation}
where $ m,n=-M,\ldots,M$ and the kinetic energy in recoil units is
\begin{equation}
 \frac{T_{mn}(\kvec)}{\ER}=
 \frac{(k_x+2\pi m)^2+(k_y+2\pi n)^2}{\pi^2}.
\end{equation}
The ordinary lattice has Fourier coefficients
\begin{equation}
 V_{0,0}=s_\gamma,
 \qquad
 V_{\pm1,0}=V_{0,\pm1}=-\frac{s_\gamma}{4}.
\end{equation}
Expanding the drive Stark term of Eq.~\eqref{eq:full_inplane_potential} gives
\begin{align}
    & -v_\gamma[2\cos(2\pi x)-2\cos(2\pi y)]^2 \nonumber\\
    &={} 
    -4v_\gamma
    -2v_\gamma\cos(4\pi x)
    \nonumber\\
    &\quad 
    -2v_\gamma\cos(4\pi y)
    \nonumber\\
    &\quad
    +8v_\gamma\cos(2\pi x)\cos(2\pi y),
\end{align}
so that the additional Fourier coefficients are
\begin{equation}
 V_{0,0}=-4v_\gamma,
 \:\:\:
 V_{\pm2,0}=V_{0,\pm2}=-v_\gamma,
 \:\:\:
 V_{\pm1,\pm1}=2v_\gamma.
\end{equation}
The results reported in the main text use $M=7$, a $12\times12$ Brillouin-zone mesh, and a real-space Wannier grid with spacing $0.025a$.

\subsection{Gauge fixing}

At each momentum, the lowest-band eigenvector is multiplied by a phase such that
\begin{equation}
 \psi_{\kvec}(0,0)=\sum_{m,n}c_{mn}(\kvec)
\end{equation}
is real and positive, and the Wannier function is taken as the normalized sum of the gauge-fixed Bloch states. 

\subsection{Vertical dimer orbitals and interactions}
\label{app:vertical}

The vertical potential of Eq.~\eqref{eq:vertical_potential} has supercell period $2a_s$. Its Bloch bands are obtained in a one-dimensional plane-wave basis with reciprocal vectors $\pi n/a_s$. For an isolated dimer, the bonding and antibonding bands are separated by approximately $2t_\perp$, so we identify
\begin{equation}
 t_\perp=\frac{1}{2}
 \avg{E_1(k_z)-E_0(k_z)}_{k_z},
\end{equation}
and calculate half the width of the lowest band as the residual hopping between neighboring dimers, which is the amplitude $t^{\gamma}_{\text{inter}}$.

Evaluating the interaction integrals requires an orbital localized in a single well rather than a dimer-centered bonding state. The two lowest eigenstates $\psi_0$ and $\psi_1$ of an isolated double well are therefore combined as
\begin{equation}
 \phi_L(z)=\frac{\psi_0(z)-\psi_1(z)}{\sqrt2}.
\end{equation}
The three-dimensional density overlaps then factorize into in-plane and vertical pieces, for example
\begin{equation}
 U_{gg}=
 \frac{4\pi\hbar^2a_{gg}}{m}
 \int dx\,dy\,|w_g(x,y)|^4
 \int dz\,|\phi_g(z)|^4,
\end{equation}
and for one $g$ and one $e$ atom
\begin{equation}
 U_{eg}^{\pm}=
 \frac{4\pi\hbar^2a_{eg}^{\pm}}{m}
 \int d^3r\,|w_g(\rvec)|^2|w_e(\rvec)|^2,
\end{equation}
from which $V$ and $\Jex$ follow through Eq.~\eqref{eq:V_Jex_definition}.

\section{Vertical lattice optimization}
\label{app:vertical_wavelength}

\begin{figure}
\includegraphics[width=0.9\columnwidth]{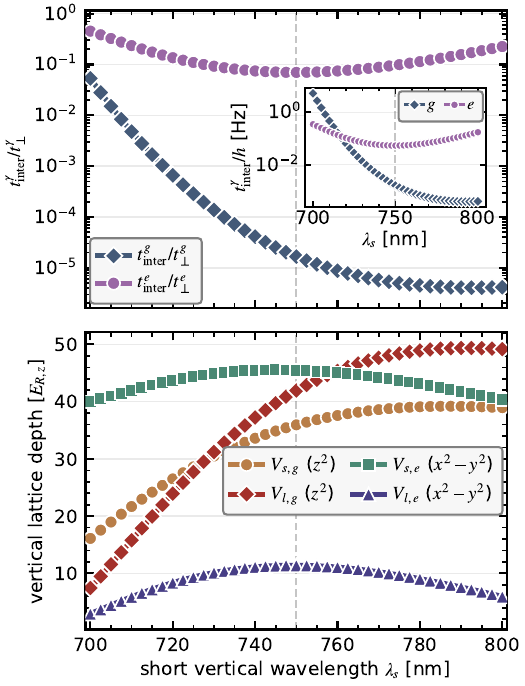}
\caption{\textbf{Vertical-lattice wavelength and bilayer isolation.}
(a) residual inter-bilayer hopping $t_{\mathrm{inter}}^\gamma$, normalized by the corresponding desired intra-bilayer hopping $t_\perp^\gamma$, as a function of the short-lattice wavelength $\lambda_s$. At each wavelength, the vertical lattice depths are reoptimized such that the target hopping ratios $t_{\perp}^{xx}/t_{\parallel}^{xx}$, $t_{\perp}^{zz}/t_{\parallel}^{xx}$ remain fixed. The inset shows the corresponding absolute inter-bilayer hoppings in Hz. (b) Optimized short- and long-lattice depths for both clock states. Vertical energies are expressed in the wavelength-dependent recoil energy $E_{R,z}(\lambda_s)=h^2/(8ma_s^2)$ with $a_s=\lambda_s/2$.}
\label{fig:vertical_wavelength_scan}
\end{figure}

In the main text, we fix the vertical superlattice wavelengths to $\lambda_s=750$~nm. We now examine how this choice affects the isolation of individual bilayers. For each value of $\lambda_s$, we reoptimize the two independent vertical lattice intensities such that the desired intra-bilayer hoppings $t_\perp^{gg}$ and $t_\perp^{ee}$ remain fixed at their target values. We then extract the residual inter-bilayer hopping $t_{\rm inter}^\gamma$ from half the width of the lowest vertical Bloch band. The resulting isolation ratios $t_{\rm inter}^\gamma/t_\perp^\gamma$ are shown in Fig.~\ref{fig:vertical_wavelength_scan}, together with the corresponding optimized lattice depths.

The strong wavelength dependence can be understood directly from the structure of the vertical potential,
\begin{equation}
V_\gamma^z(z)
=
V_{s,\gamma}\cos^2\!\left(\frac{\pi z}{a_s}\right)
+
V_{l,\gamma}\sin^2\!\left(\frac{\pi z}{2a_s}\right),
\label{eq:vertical_potential_appendix}
\end{equation}
with $a_s=\lambda_s/2$. Introducing
\begin{equation}
q_\gamma=\frac{V_{l,\gamma}}{4V_{s,\gamma}},
\end{equation}
the barriers separating the two wells within one bilayer and neighboring bilayers are, respectively,
\begin{equation}
\begin{aligned}
&\Delta V_{\rm intra}^{\gamma} = V_{s,\gamma}(1-q_\gamma)^2, \\
&\Delta V_{\rm inter}^{\gamma} = V_{s,\gamma}(1+q_\gamma)^2.
\label{eq:vertical_barriers}
\end{aligned}
\end{equation}
Thus, increasing the long-lattice contribution has the useful dual effect of lowering the barrier within a bilayer while simultaneously raising the barrier between adjacent bilayers.

The two clock states experience different superlattice shapes because the short- and long-wavelength components have different state-dependent polarizabilities. Defining
\begin{equation}
r_s=
\frac{\alpha_e(\lambda_s)}{\alpha_g(\lambda_s)},
\qquad
r_l=
\frac{\alpha_e(2\lambda_s)}{\alpha_g(2\lambda_s)},
\end{equation}
one has
\begin{equation}
V_{s,e}=r_s V_{s,g},
\qquad
V_{l,e}=r_l V_{l,g},
\qquad
\frac{q_e}{q_g}=\frac{r_l}{r_s}.
\label{eq:q_ratio}
\end{equation}
A strong difference between $r_l$ and $r_s$ thus allows the two states to experience substantially different double-well geometries even when their short-lattice depths are comparable. In particular, a small $r_l/r_s$ permits the $g$ state to acquire a large intra-bilayer tunneling while retaining a high barrier toward neighboring bilayers, whereas the $e$ state remains much more weakly coupled across the dimer.

Fig.~\ref{fig:vertical_wavelength_scan}~(a) shows the double well leakage as a function of the (short) lattice wavelength. Increasing $\lambda_s$ strongly suppresses the residual $g$-state inter-bilayer tunneling, while the $e$-state leakage exhibits a broader minimum and eventually increases again. The optimized lattice depths vary smoothly across the scan, shown in Fig.~\ref{fig:vertical_wavelength_scan}~(b). The chosen wavelength in the main text therefore represents a compromise between reproducing the required intra-bilayer hopping hierarchy and simultaneously minimizing unwanted tunneling between neighboring bilayers.

\section{Alternative orbital assignment}
\label{app:alternative_mapping}
\begin{figure}
\includegraphics[width=0.98\columnwidth]{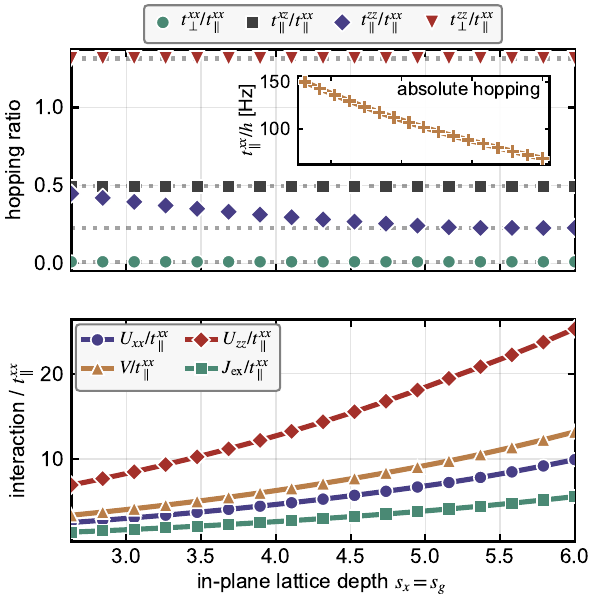}
\caption{\textbf{Alternative orbital assignment.} (a) Hopping ratios normalized to $t^{xx}_{\parallel}=t^{gg}_{\parallel}$, which can be pinned to the targets of Tab.~\ref{tab:hoppings} for lattice depths $s_g \gtrsim 5 E_R$. (b) On-site interactions, showing that the combination of the $e$-state scattering length with the stronger localization of the $z$ orbital produces a substantially larger separation between $U_x$ and $U_z$ than in Fig.~\ref{fig:optimization_summary}.}
\label{fig:original_scan}
\end{figure}

For comparison, we analyze the alternative assignment
\begin{equation}
 g={}^1S_0\leftrightarrow x=\dxy,
 \qquad
 e={}^3P_0\leftrightarrow z=\dz.
 \label{eq:original_mapping}
\end{equation}
At an in-plane wavelength near $700$~nm one finds $\alpha_e/\alpha_g\simeq2.4$, so the $e$ state is trapped more deeply and $t^{gg}_{\parallel}>t^{ee}_{\parallel}$, as required here. The vertical hierarchy can be generated with $550$ and $1100$~nm superlattice components. Fig.~\ref{fig:original_scan} shows that the hopping ratios of Tab.~\ref{tab:hoppings} can indeed be pinned across a range of lattice depths.

However, its interaction and lattice-depth costs are less favorable. Because the $z$ orbital is encoded in the $e$ state, the larger scattering length $a_{ee}$ is combined with the more localized in-plane Wannier function, and both effects increase $U_z$. Similarly, the $x$ orbital is encoded in $g$, so the smaller $a_{gg}$ is combined with the broader Wannier function, and both effects decrease $U_x$. The two intra-orbital repulsions therefore move apart rather than compensating, which is the opposite of the behavior of that required in the bilayer nickelate model. Furthermore, as shown in the upper panel of Fig.~\ref{fig:original_scan}, the in-plane hopping ratio is limited in programmability, deviating from the DFT target below in-plane lattice depths of around $s_g \lesssim 5 E_R$. The vertical superlattice is also substantially more demanding in terms of lattice depths, requiring $V_{s,g},V_{l,g} \sim 200\,\ERz$.

\section{ Floquet-engineering of spin-independent inter-orbital tunneling \label{app:pulse_engineered_tunneling} }
Engineering an inter-orbital and inter-site hopping term requires us to drive the $^{1}S_0$ and $^{3}P_0$ clock states of $^{87}$Sr on two hyperfine magnetic sublevels simultaneously. The catch is that we want to drive both these sublevels with the same Rabi frequency. For concreteness, we consider the encoding of spins into the hyperfine sublevels $\uparrow \equiv (m_F = 9/2)$ and $\downarrow \equiv (m_F=7/2)$, so that the two driven transitions are:
\begin{equation}
\begin{aligned}
    & |g; \downarrow\rangle
    \underset{\pi}{\:\:\rightarrow\:\:} 
    |e; \downarrow\rangle
    \\
    & |g; \uparrow\rangle
    \underset{\pi}{\:\:\rightarrow\:\:} 
    |e; \uparrow\rangle.
\end{aligned}
\end{equation}
All laser polarizations, labeled under the state-transition arrow, are with respect to a small bias magnetic field ($\sim 1 - 5$ G) that sets the quantization axis.   
Due to the Clebsch-Gordan coefficients, driving these transitions with a single-tone laser will result in different Rabi frequencies since $\Omega_{m_F} \propto m_F$. 
To fix this discrepancy, we consider dressing the two hyperfine sublevels so that they become strongly mixed. One might consider doing so directly with a radio-frequency (RF) drive, although the small nuclear spin magnetic dipole moment ($\mu \approx 6 \times 10^{-4} \mu_B$) makes this technically demanding. Instead, we consider driving this transition through a two-photon Raman process. 

\subsection{Equalizing clock Rabi rates}

For a Raman drive, we will require two lasers per electronic clock state:
\begin{equation}
\begin{aligned}
    & |g; \uparrow\rangle 
    \underset{\pi}{\:\:\rightarrow\:\:}
    | ^{3}P_1; F=9/2, m_F=9/2 \rangle 
    \underset{\sigma^+}{\:\:\leftarrow\:\:}
    |g; \downarrow\rangle,
    \\
    & |e; \uparrow\rangle
    \underset{\pi}{\:\:\rightarrow\:\:}
    | ^{3}D_1; F=9/2, m_F=9/2 \rangle 
    \underset{\sigma^+}{\:\:\leftarrow\:\:}
    |e; \downarrow\rangle,
\end{aligned}
\end{equation}
one of which is $\pi$-polarized and the other $\sigma^+$ polarized. If both the Raman lasers are far detuned from the intermediate state, we can adiabatically eliminate it from the three-level system
\begin{align}
    \boldsymbol{{\cal H}}_{\rm Raman}
    &=
    \hbar
    \begin{pmatrix}
        0 & \Omega_{\pi}/2 & 0 \\
        \Omega_{\pi}/2 & \Delta_{\pi} & \Omega_{\sigma^{+}}/2 \\
        0 & \Omega_{\sigma^{+}}/2 & \delta_R
    \end{pmatrix},
\end{align}
with two-photon detuning $\delta_R$ and $\pi$-laser detuning $\Delta_{\pi}$,  
to leave an effective two-level system of the two relevant hyperfine sublevels
\begin{align}
    \boldsymbol{{\cal H}}_{\rm Raman}
    &\approx 
    \hbar
    \begin{pmatrix}
        0 & \tilde{\Omega}/2 \\
        \tilde{\Omega}/2 & \tilde{\Delta}
    \end{pmatrix},
\end{align}
where $\tilde{\Delta} = \delta_R + (\Omega_{\pi}^2 - \Omega_{\sigma^{+}}^2) / (4\Delta_{\pi})$ is the effective detuning and $\tilde{\Omega} = \Omega_{\pi} \Omega_{\sigma^{+}} / (2\Delta_{\pi})$ the effective Rabi frequency. Both Hamiltonians above are written in the rotating frame.

We can tune the strength and detuning of the two Raman lasers such that $\tilde{\Delta} = 0$, 
which creates a perfect equal superposition of the two hyperfine sublevels with opposite parity $|\pm\rangle = (|F=9/2, m_F=9/2\rangle \pm |F=9/2, m_F=7/2\rangle) / \sqrt{ 2 }$. These states have corresponding energies $E_{\pm} = \pm \hbar\tilde{\Omega} / 2$. If we now drive the clock transition (with a $\pi$-polarized laser), the effective Rabi frequency becomes
\begin{align}
    \Omega_{\rm avg}
    &=
    \frac{ \Omega_{9/2} + \Omega_{7/2} }{ 2 },
\end{align}
for both odd and even parity states. As mentioned in the main text, we achieve the above effect with a pulsed sequence instead of continuous lasers to mitigate photon scatter and loss. Doing so still requires isolation of the spin-encoded pair, an issue we now turn to addressing. 

\subsection{Isolating spin-1/2 states}

Because all the hyperfine sublevels are equally split in the presence of a small magnetic field, we will need to isolate the $m_F = 9/2$ and $7/2$ states from the other sublevels. To do so, we utilize the technique in Ref.~\cite{Barnes2022} that resonantly dresses the closest $m_F = 5/2$ hyperfine sublevel and shifts it away. 
We choose to dress the $m_F=5/2$ states in the ground and excited clock states with dipole-coupled ``blocker" states:  
\begin{equation}
\begin{aligned}
    | b_g \rangle
    =
    |^{3}P_1; F=7/2, m_F=7/2\rangle, 
    \\
    | b_e \rangle
    =
    |^{3}D_1; F=7/2, m_F=7/2\rangle,
\end{aligned}
\end{equation}
respectively. The dressing is achieved with $\sigma^+$-polarized light, suppressing leakage into the $m_F=5/2$ states. 

\subsection{Spin-flip control sequence} 

Finally, we turn to the Floquet-engineering of the Raman-driven Hamiltonian, where we propose the symmetric pulse sequence 
\begin{align} \label{eq:sequence-short}
  \boxed{\frac{T}{4}}
  \;\longrightarrow\;
  \boxed{S}
  \;\longrightarrow\;
  \boxed{\frac{T}{2}}
  \;\longrightarrow\;
  \boxed{S^{-1}}
  \;\longrightarrow\;
  \boxed{\frac{T}{4}},
\end{align}
where the clock laser acts during the three clock-drive segments based on period $T$, and $S$ is a $m_F=9/2\leftrightarrow7/2$ Raman $\pi$-pulse applied simultaneously to both the $^1S_0$ and $^3P_0$ electronic manifolds. The second spin-flip is implemented about the opposite Raman rotation axis (a rotary-echo phase shift of $\pi$) to suppress systematic pulse-area errors.
The clock driving laser is turned off during both spin-flip pulses, leaving only the Raman and blocker lasers active during the spin-flip segments. The converse is true in the clock-drive segments. Each spin state spends a time of $T/2$ in each nuclear-spin state for a period $T$. If the clock driving lasers are exactly resonant and rotate about the same Bloch-sphere axis, then driving the clock transition on either spin state will result in Hamiltonians that mutually commute. 

We note that using a small bias magnetic field $B$ to define the quantization axis, the $^{1}S_0$ Zeeman splitting is approximately $(184.4 \times B)$ Hz/$m_F$/G. The two bare clock lines are therefore not degenerate, and a single $698$-nm carrier is placed at their midpoint. Their opposite detunings do not commute with the clock-drive term, so the symmetric spin-flip sequence suppresses but does not completely eliminate higher-order errors. The ground and excited Raman phases must also be referenced to the clock coherence. For a spin-flip beginning at time $t$, we impose the phase condition $\phi_e-\phi_g = \Delta\omega_c t + \phi_0$, where $\Delta\omega_c$ is the frequency separation of the adjacent bare clock lines and $\phi_0$ accounts for all static phase offsets that can be calibrated away. This condition can be enforced by deterministic phase resets since we know the Zeeman splitting.

The Floquet Hamiltonian engineered by this pulse sequence can be obtained using the Magnus expansion. The zeroth-order average Hamiltonian is given by $\tilde{H}_{0}=\sum_i H_i \tau_i$, where $H_i$ is the toggling-frame Hamiltonian during the $i$-th interval, and $\tau_i$ is the corresponding duration. For our proposed Floquet pulse sequence, the mirror symmetry of the sequence cancels all odd-order Magnus terms. Furthermore, under the condition that $T$ is much smaller than the inverse energy scales of the system's Hamiltonian, the remaining higher-order terms are negligible. Consequently, the Floquet Hamiltonian is well approximated by the zeroth-order average Hamiltonian, $H_{\rm F} \approx \tilde{H}_{0}$.

\bibliographystyle{apsrev4-2}
\bibliography{ref}

\end{document}